\documentclass{emulateapj}

\usepackage{graphicx}
\usepackage{amssymb}
\usepackage{longtable}
\usepackage{epsf}
\usepackage{apjfonts}
\usepackage{rotating}

\slugcomment{Accepted to Astrophysical Journal}

\shorttitle{Confirmation of monster compact galaxy at z=3.35}
\shortauthors{Marsan et al.}

\begin{document}


\title{Spectroscopic Confirmation of an Ultra Massive and Compact Galaxy at 
$z$=3.35:\\A Detailed Look at an Early Progenitor of Local Most Massive 
Ellipticals}

\author{Z.~Cemile~Marsan\altaffilmark{1}, Danilo~Marchesini\altaffilmark{1}, 
Gabriel B. Brammer\altaffilmark{2}, Mauro Stefanon\altaffilmark{3}, 
Adam Muzzin\altaffilmark{4}, Alberto Fern\'andez-Soto\altaffilmark{5,6},
Stefan Geier\altaffilmark{7},
Kevin N. Hainline\altaffilmark{8},
Huib Intema\altaffilmark{9}
Alexander Karim\altaffilmark{10}, 
Ivo Labb\'e\altaffilmark{4}, Sune Toft\altaffilmark{11}, Pieter G. van 
Dokkum\altaffilmark{12}}

\altaffiltext{1}{Department of Physics and Astronomy, Tufts University, 
Medford, MA 02155, USA}
\altaffiltext{2}{Space Telescope Science Institute, 3700 San Martin Drive, 
Baltimore, MD 21218, USA}
\altaffiltext{3}{Physics and Astronomy Department, University of Missouri, 
Columbia, MO 65211, USA}
\altaffiltext{4}{Leiden Observatory, Leiden University, PO Box 9513, NL-2300 
RA Leiden, The Netherlands}
\altaffiltext{5}{Instituto de F\'{\i}sica de Cantabria (CSIC-Universidad de 
Cantabria), Avda. de los Castros s/n, 39005-Santander, Spain}
\altaffiltext{6}{Unidad Asociada Observatori Astron\`omic (IFCA-Universitat de 
Val\`encia), C. Catedr\'atico Jos\'e Beltr\'an 2, 46980-Paterna, Spain}
\altaffiltext{7}{Instituto de Astrof\'{\i}sica de Canarias (IAC), E-38205 La Laguna, Tenerife, Spain}
\altaffiltext{8}{Department of Physics and Astronomy, Dartmouth College, 
Hanover, NH 03755, USA}
\altaffiltext{9}{National Radio Astronomy Observatory, 1003 Lopezville Road, Socorro, NM 87801-0387, USA}
\altaffiltext{10}{Argelander-Institut f\"ur Astronomie, Universit\"at Bonn, Auf dem H\"ugel 71, D-53121 Bonn, Germany}
\altaffiltext{11}{Dark Cosmology Centre, Niels Bohr Institute, University of 
Copenhagen, Juliane Maries Vej 30, DK-2100 Copenhagen, Denmark}
\altaffiltext{12}{Department of Astronomy, Yale University, New Haven, CT 
06511, USA}

\begin{abstract}
We present the first spectroscopic confirmation of an ultra-massive galaxy at redshift $z>3$ using data from Keck-NIRSPEC, VLT-X-shooter, and GTC-Osiris. We detect strong [OIII] and Ly$\alpha$ emission, and weak [OII], CIV, and HeII, placing C1-23152 at a spectroscopic redshift of $z_{spec}=3.351$. The modeling of the emission-line corrected spectral energy distribution results in a best-fit stellar mass of $M_{*}=3.1^{+0.6}_{-0.7}\times10^{11}$ M$_{\sun}$, a star-formation rate of $<$7~M$_{\sun}$~yr$^{-1}$, and negligible dust extinction. The stars appear to have formed in a short intense burst $\sim$300-500 Myr prior to the observation epoch, setting the formation redshift of this galaxy at z$\sim$4.1. From the analysis of the line ratios and widths, and the observed flux at 24$\mu$m, we confirm the presence of a luminous hidden active galactic nucleus (AGN), with bolometric luminosity of $\sim 10^{46}$ erg s$^{-1}$. Potential contamination of the observed SED from the AGN continuum is constrained, placing a lower limit on the stellar mass of $2~\times~10^{11}$~M$_{\sun}$. HST/WFC3 H$_{\rm 160}$ and ACS I$_{\rm 814}$ images are modeled, resulting in an effective radius of $r_{\rm e} \sim 1$ kpc in the H$_{\rm 160}$ band and a S\'ersic index $n \sim 4.4$. This object may be a prototype of the progenitors of local most massive elliptical galaxies in the first 2 Gyr of cosmic history, having formed most of its stars at $z>4$ in a highly dissipative, intense, and short burst of star formation. C1-23152 is completing its transition to a post-starburst phase while hosting a powerful AGN, potentially responsible for the quenching of the star formation activity.
\end{abstract}



\section{INTRODUCTION}\label{sec-in}

One of the most fundamental questions regarding galaxy formation and evolution 
is when and how the most massive galaxies in the universe formed. In the 
current $\Lambda$CDM paradigm, the dominant structures in the universe are 
dark matter haloes that grow out of primordial density perturbations through 
gravitational collapse \citep{whiterees78}. Simulations and analytical models 
establish that this process proceeds in a hierarchical, bottom-up manner with 
low mass haloes forming first and subsequently growing by continued accretion 
and merging to form more massive haloes at later times 
\citep{whitefrenk91, kauffmannwhite93, kauffmann99}. In contrast, 
observational studies suggest that the stellar baryonic component of haloes 
assemble in an anti-hierarchical, top-down manner \citep{cowie96, fontanot09}.

Closely related to this issue is the intriguing finding that the number density 
of the most massive galaxies ($M_{*} > 3\times10^{11}M_{\sun}$) seems to 
evolve very little from $z \sim 4$ to $z \sim 1.5$ 
\citep{perezgonzalez08,marchesini09,marchesini10,muzzin13b}, suggesting that 
very massive galaxies were already in place at $z \sim 3.5$, and implying that 
their stellar content was assembled rapidly in the first $\sim$1.5 Gyr of 
cosmic history. In contrast to the homogeneous population of massive galaxies in the nearby universe, 
the population of massive galaxies in the early universe is found to 
span a diverse range in stellar ages, star formation rates, and dust contents 
\citep{marchesini10, ilbert13, muzzin13b, stefanon13, spitler14, marchesini14}. Studies of
stellar mass complete samples of galaxies at high redshift reveal populations 
of both evolved quiescent galaxies and star-forming galaxies with varying
degrees of dust attenuation \citep{marchesini10,stefanon13, spitler14}. An additional class
of dust-enshrouded and rigorously star forming high-$z$ massive galaxies are uncovered 
through detections at longer wavelengths (sub-millimeter galaxies, SMGs; \citealt{smail02}). 
The SMG population is dominated by galaxies with SFRs $\sim10\times$ 
more intense than main sequence galaxies at these redshifts with indications of ongoing merging 
\citep{tacconi06,ivison13, fu13}. 

Theoretical models of galaxy formation and evolution severely 
under-predict the observed number density of 
galaxies at $3<z<4$ with $\log{(M_{*}/M_{\sun})}>11.5$ \citep{fontanot09, 
marchesini09, marchesini10}, but systematic uncertainties in both photometric redshift and stellar 
mass estimates remain large.
While several massive SMGs have spectroscopically confirmed redshifts at $z>3$ (e.g., \citealt{schinnerer08, smolcic11, 
walter12, toft14}), they suffer from significant 
uncertainties in derived stellar population parameters \citep{michalowski12}. Spectroscopically 
confirming the redshift of rest-frame optically selected galaxies is challenging as these sources 
are faint in the observers' optical, requiring significant exposure times at
even the largest telescopes (but see \citealt{strazzullo13} for a massive $z_{spec}=2.99$ passive galaxy).  

Studies find that, of the massive galaxy population (e.g., $M_{*} > 10^{11} M_{\sun}$) at $z \sim 2$, 
approximately $40\%$ are no longer forming stars \citep{whitaker11, brammer11, 
muzzin13b}, and this population of quiescent galaxies undergoes significant 
size evolution. Specifically, massive quiescent galaxies are found to 
be compact, particularly those with the lowest star formation rates, with 
sizes a factor of $\sim$4-5x smaller at $z\sim2$ compared to local quiescent 
galaxies with similar stellar masses (\citealt{daddi05, trujillo06, toft07, 
trujillo07, zirm07, cimatti08, vandokkum08, franx08, vanderwel08, bezanson09, 
vandokkum10}; see \citealt{vanderwel14} for the latest comprehensive analysis 
on the size evolution of quiescent and star-forming galaxies). It has been 
suggested that dry minor mergers can cause a considerable growth in size 
without significantly increasing the mass of the galaxy, whereas major mergers 
tend to increase the mass without increasing the size much 
\citep{naab06,trujillo11}. A consistent size evolution through minor merging 
can be achieved as less massive satellite galaxies deposit mass preferentially 
in the outer radii of more massive central galaxies (commonly referred to as 
{\it inside-out growth}). Several recent works have indeed shown that the mass 
at fixed inner physical radius of massive, quiescent galaxies is nearly 
constant, whereas the mass in the outer regions undergoes a significant 
increase (a factor of $\sim$4 since $z\sim$2), supporting the inside-out 
growth of massive, quiescent galaxies (e.g., 
\citealt{vandokkum10, vandesande13, patel13}).

In order to understand the onset and evolution of massive galaxies, several 
recent works have aimed at identifying the potential progenitors of compact, 
massive quiescent galaxies observed at $z\sim$2, based on number density 
considerations \citep{barro13, stefanon13, patel13, williams14}. Two 
evolutionary tracks have been proposed for the formation of massive, quiescent 
galaxies: an early ($z > 2$) formation path of rapidly quenched compact 
star-forming galaxies fading into compact, quiescent galaxies that later 
enlarge within the quiescent phase; and a late-arrival ($z < 2$) path in 
which quiescent galaxies are formed through the quenching of more extended 
star-forming galaxies with lower mass densities \citep{barro13}. 

Most recently, \citet{marchesini14} studied the evolution of the properties of 
the progenitors of ultra-massive galaxies (log($M_{*}/M_{\sun}) \approx$ 11.8; 
UMGs) at $z \sim 0$ with a semi-empirical approach using abundance matching. 
At $ 2 < z < 3$, the progenitors of local UMGs are found to be dominated by 
massive ($M_{*}\approx 2\times 10^{11} M_{\sun}$), dusty ($A_{\rm V }\sim$ 1-2.2 
mag), and star forming (SFR$\sim$100-400 $\rm{M_{\sun}yr^{-1}}$) galaxies, 
although a small fraction ($\sim15\%$) of them are found to be already 
quiescent at $2.5<z<3.0$. The $2.5<z<3.0$ quiescent progenitors of local UMGs 
are found to have properties typical of young (0.6-1 Gyr) 
post-starburst galaxies with little dust extinction ($A_{\rm V}\sim$0.4 mag) 
and strong Balmer breaks, showing a large scatter in rest-frame $U-V$ color 
($\sim$0.2 mag). The existence of such quiescent progenitors at $z$ = 2.75 
indicates that the early assembly of the massive end of the local quiescent 
red-sequence must have started at $z>3$, as also implied by archeological 
studies in the local universe (e.g., \citealt{thomas05, gallazzi06, thomas10}).

Evidence for the existence of massive ($M_{*}>10^{11}$) galaxies at $z$ >3 with 
suppressed star formation has been found through studies of deep near-infrared 
selected catalogs \citep{wiklind08, mancini09, marchesini10, stefanon13, 
muzzin13b, straatman14}, with accurate photometric redshifts and well-sampled 
SEDs. However, spectroscopic observations are necessary to confirm 
the redshifts and to better characterize the properties of these galaxies. It 
is also of vital importance to determine AGN and/or starburst contamination to 
the optical-to-MIR SEDs of these sources to properly constrain the high-mass 
end of the stellar mass function at these redshifts. 

In this paper, we present the first spectroscopic confirmation of a very 
massive galaxy at $z>3$, C1-23152, with a detailed investigation of its 
stellar population and structural properties, as well as the AGN energetics. 
C1-23152 is the brightest ($K=20.3$~mag, AB) galaxy within the stellar mass 
complete ($M_{*}>2.5\times 10^{11}$~M$_{\odot}$) sample of 14 galaxies at 
$3<z<4$ presented in \citet{marchesini10} was initially selected from the 
NEWFIRM Medium-Band Survey (NMBS; \citealt{whitaker11}). C1-23152 is in the 
COSMOS field \citep{scoville07}, and has a photometric redshift 
$z_{\rm phot}=3.29\pm0.06$ and a stellar mass of 
$\log{(M_{*}/M_{\odot})}=11.42$ \citep{marchesini10}. By combining the broad- 
and medium-band photometry from NMBS with optical and near-infrared (NIR) 
spectroscopic data from both ground- and space-based facilities, we 
spectroscopically confirm the very large stellar mass of C1-23152 through 
detailed modeling of its spectral energy distribution (SED), and provide robust 
evidence for the evolutionary path linking C1-23152 to local giant elliptical 
galaxies. 

This paper is organized as follows: In \S\ref{sec-ss}, we present the 
observations of C1-23152, including extensive space- and ground-based 
spectroscopy from the ultra-violet (UV) to the near-infrared (NIR), as well as 
space-based imaging. In \S\ref{sec-analysis}, we present the analysis of the 
spectra and the modeling of the SED, and the derivation of the structural 
parameters from space-based imaging. The results are summarized and discussed 
in \S\ref{sec-disc}. We assume $\Omega_{\rm M}=0.3$, $\Omega_{\rm \Lambda}=0.7$, 
$H_{\rm 0}=70$~km~s$^{-1}$~Mpc$^{-1}$, and a \citet{kroupa01} initial mass 
function (IMF) throughout the paper. All magnitudes are on the AB system. 


\section{Data}\label{sec-ss}

\subsection{Broad- and medium-band photometry from the NMBS}\label{subsec-zphot}

The target C1-23152 (RA=10$^{\rm h}$00$^{\rm m}$27$^{\rm s}$.81, 
DEC=+02$^{\rm d}$33$^{\rm m}$ 49$^{\rm s}$.3; J2000) is selected from the 
mass-complete ($M_{*}>2.5\times10^{11}$~M$_{\sun}$) sample of galaxies at 
$3<z<4$ constructed from the NMBS and presented in \citet{marchesini10}. 
C1-23152 lies in the COSMOS field \citep{scoville07}, with excellent 
supporting data. The NMBS photometry presented in \citet{whitaker11} includes 
deep optical \emph{ugriz} data from the CFHT Legacy Survey, deep 
\emph{Spitzer}-IRAC and MIPS imaging \citep{sanders07}, deep Subaru images 
with the $B_{J}V_{J}r^{+}i^{+}z^{+}$ broadband filters \citep{capak07}, Subaru 
images with 12 optical intermediate-band filters from 427 to 827~nm 
\citep{taniguchi07}, $JHK_{\rm S}$ broad-band imaging from the WIRCam Deep 
Survey (WIRDS; \citealt{mccracken10}), \emph{Galaxy Evolution 
Explorer} ({\it GALEX}) photometry in the FUV (150 nm) and NUV (225 nm) 
passbands \citep{martin05}, and NIR imaging with NEWFIRM using the five 
medium-band filters $J_{\rm 1}$, $J_{\rm 2}$, $J_{\rm 3}$, $H_{\rm 1}$, 
$H_{\rm 2}$, and broad-band $K$ from NMBS \citep{vandokkum09}, for a total of 
37 photometric points. In the publicly released v5.1 NMBS catalog, C1-23152 is 
identified as C1-35502, and has total apparent magnitudes of 
$K_{\rm S}=20.31$~mag and $r=23.24$~mag.


\subsection{Ground-based spectroscopy and data reduction}\label{subsec-sed}

In this section we describe the ground-based spectroscopic data of C1-23152 
obtained from Keck-NIRSPEC, VLT-X-shooter, and GTC-Osiris as part of follow-up 
programs aimed at spectroscopically confirming the existence of very massive 
galaxies at $z>3$. Along with the observational techniques, we provide a 
summary of the data reduction. The detailed description of the data reduction 
and the full sample will be presented elsewhere (C. Marsan et al. 2014; in 
preparation). Table~\ref{tab-obs} summarizes the wavelength coverage, total 
exposure time, seeing, date of observations, and the slit width of the 
spectroscopic observations. 

\begin{deluxetable}{lccccc}
\centering
\tablecaption{Ground-based Spectroscopic Observations\label{tab-obs}}
\tablehead{\colhead{Band/Arm} & \colhead{$\lambda_{\rm range}$} & 
\colhead{$t_{\rm exp}$} & \colhead{FWHM} & \colhead{Date} & \colhead{Width}\\
 &  ($\mu$) &  ($min$) & ($^{\prime \prime}$) &  & ($^{\prime \prime}$)} 
\startdata
\multicolumn{6}{l}{\it Keck-NIRSPEC} \\
N5 (H)  & 1.48 - 1.71 & 75 & 0.7 & 11 Feb 2011  & 0.72 \\
N7 (K)  & 1.94 - 2.37 & 75 & 0.7 & 11 Feb 2011 & 0.72 \\
\multicolumn{6}{l}{\it VLT-X-shooter}\\
UVB & 0.3 - 0.59  & 35 & 1.16 & 27 May 2011 & 1.0 \\
VIS & 0.53 - 1.02 & 35 & 1.16 & 27 May 2011 & 0.9 \\
NIR & 0.99 - 2.48 & 40 & 1.16 & 27 May 2011 & 0.9 \\
\multicolumn{6}{l}{\it GTC-Osiris}\\
R1000B & 0.37 - 0.77  & 42 & 0.9 & Nov 2011 & 0.8 \\
R1000B & 0.37 - 0.77  & 42 & 1.6 & Feb 2012 & 0.8 \\
R1000B & 0.37 - 0.77  & 45 & 1.0 & Jan 2013 & 0.8 \\

\enddata
\tablecomments{$\lambda_{\rm range}$ is the wavelength range covered by the 
instrumental setup; $t_{\rm exp}$ is the on-source exposure time in minutes; 
FWHM is the average seeing in arcsec of the observations. The last two columns 
list the nights during which the spectroscopic data were taken and the width of 
the used slit in arcsec, respectively.}
\end{deluxetable}

\subsubsection{Keck-NIRSPEC}
We observed C1-23152 using Keck-NIRSPEC \citep{mclean98} in cross-dispersed mode and the 
42$^{\prime \prime}$$\times$0.76$^{\prime \prime}$ slit for both N5 (H) and N7 (K) 
filters, obtaining a wavelength coverage of 1.48-1.71 $\mu$m and 1.94-2.37 
$\mu$m, and a spectral resolution of R=1100 and R=1500, respectively. The 
observations were carried out on the night of February 11, 2011 as part of the 
NOAO program 2011A-0514 (PI: Marchesini) with a typical seeing of 
0.7$^{\prime \prime}$ which worsened throughout the sequence of observations 
due to cloudy variable weather. Observations were conducted following an 
ABA$^{\prime}$B$^{\prime}$ on-source dither pattern. The orientation of the slit 
was set to include a bright point source to serve as reference when analyzing 
and combining the two dimensional rectified frames. The target was acquired 
using blind offsets from a nearby bright star. The alignment of the offset 
star in the slit was checked before each individual 900 sec exposure and 
corrected when necessary. Before and after the observing sequence, a 
spectrophotometric standard and an AV0 star was observed for the purpose of 
correcting for telluric absorption and detector response. 

The data reduction for the NIRSPEC observations used a combination of custom 
IDL scripts  and standard IRAF tasks\footnote{IRAF is distributed by the 
National Optical Astronomy Observatory, which is operated by the
Association of Universities for Research in Astronomy (AURA), Inc., 
under cooperative agreement with the National Science Foundation.}. As the initial step, a bad pixel mask 
was created by flagging outlier pixels in dark and flat frames. The cosmic 
rays on the science frames were removed using L.A.Cosmic \citep{vandokkum01}. 
The frames and their corresponding sky spectra were rotated such that the sky 
lines are along columns, using a polynomial to interpolate between adjacent 
pixels. The spectra were wavelength calibrated by fitting Gaussian profiles 
to the OH lines in the 2D sky spectra. Each spectrum was sky subtracted using 
an adjacent spectrum with the IDL routines written by George Becker. The sky 
subtracted frames and the sky spectra were rectified to a linear wavelength 
scale. The standard star frames used to correct for atmospheric absorption and 
detector response were rectified and reduced in the same manner as the science 
frames. A one-dimensional spectrum was extracted for each telluric standard 
star (before and after science observations) by summing all the rows (along 
the spatial direction) with a flux greater than 0.1 times that of the central 
row. The average of the one-dimensional telluric star spectra was used to 
correct the two dimensional science and spectrophotometric star frames for 
telluric absorption. The one-dimensional spectrum of the spectrophotometric 
star was extracted in the same manner, and used to create a response function 
for flux calibration. The two dimensional rectified and reduced science frames 
were combined by weighting according to their signal-to-noise ratio (S/N).

\subsubsection{VLT-X-shooter}
X-shooter is a single-object, medium-resolution echelle spectrograph with 
simultaneous coverage of wavelength range 0.3-2.5$\mu$m in three arms (UVB, 
VIS, NIR) \citep{dodorico06}. The observations of our target were carried out 
in queue mode as part of the ESO program 087.A-0514 (PI: Brammer) under poor 
seeing conditions (FWHM$\sim$1.2$^{\prime \prime}$), following an 
ABA$^{\prime}$B$^{\prime}$ on-source dither pattern using the 
11$^{\prime \prime}$$\times$1.0$^{\prime \prime}$ and 
11$^{\prime \prime}$$\times$0.9$^{\prime \prime}$ slits for the UVB and VIS/NIR 
arms, respectively. This instrumental setup resulted in a spectral resolution 
of R=4200, 8250, and 4000 for the NIR, VIS and UVB arms, respectively. For calibration
purposes, telluric and spectrophotometric standard stars were observed in the same setup 
as science observations.

The data reduction for the X-shooter observations used custom scripts based on 
the standard X-shooter reduction pipeline \citep{modigliani10}. The calibration 
steps (master darks, order prediction, flat fields, and the two-dimensional 
maps for later rectification of the spectra) were run with the default 
parameters in the pipeline \citep{goldoni11}. After these five calibration 
steps, the echelle spectra were dark-subtracted, flat-fielded, and rectified, 
and the orders stitched (12, 15 and 16 for the UVB, VIS and NIR arms, 
respectively). The sky was then subtracted using adjacent exposures. Standard 
star observations were reduced with the same calibration data as the science 
frames and used to correct for  telluric absorption and  detector response. 
A final spectrum was created by mean stacking all exposures. We refer to 
\citet{geier13} for a detailed description of reduction steps of X-shooter 
spectra. 

\subsubsection{GTC-Osiris}
The target was also observed using the Gran Telescopio Canarias (GTC) with 
the Osiris optical spectrograph \citep{cepa03} in long-slit mode with the R1000B grism 
(dispersion of 2.12~\AA~pxl$^{-1}$) and a 0.8$^{\prime \prime}$ wide slit, 
resulting in a spectral resolution of $R\sim800$. Observations were carried 
out in queue mode under clear sky conditions under programs GTC8-11B and 
GTC22-12B (PI: Fern\'andez-Soto). During the 2011/2012 winter we obtained two
separate runs. The first one was acceptable (average seeing of 
0.9$^{\prime \prime}$), but the second was below the expected quality and the 
data were not useful (seeing worse than 1.5$^{\prime \prime}$). In each of them 
three exposures of 840 seconds each were taken, using offsets of 
$+$1.78$^{\prime \prime}$/0$^{\prime \prime}$/$-$1.78$^{\prime \prime}$ along the 
slit. A third run in January 2013 obtained three more exposures of 900 
seconds each, with an adequate seeing (1.0$^{\prime \prime}$) and using the same
observing strategy. In all cases the orientation of the slit was set to 
include a bright point source to serve as reference when analyzing and 
combining the two dimensional rectified frames.

Data reduction was performed using the standard long-slit package in IRAF 
\citep{valdes92}, and all previous image calibration steps (bias subtraction, 
trimming, flat/fielding) before the final frame combination for each separate 
run were also performed within IRAF. Wavelength calibration was obtained using 
the HgAr and Ne arclamp exposures provided by the Observatory, and the images 
were rectified to correct for obvious flexures along the spatial direction. An
approximate flux calibration was obtained via the observations of the
spectrophotometric standard stars G191-B2B or GD158-100.

\subsubsection{Extraction of One-dimensional Spectra}
Following \citet{horne86}, one-dimensional spectra were extracted by summing 
all adjacent lines (along the spatial direction) using weights corresponding 
to a Gaussian centered on the central row with a full width at half maximum 
(FWHM) equal to the slit width used in each observation. To correct for slit 
losses and obtain an absolute flux calibration, spectroscopic broad/medium-band 
fluxes were obtained by integrating over the corresponding filter curves, and 
a constant scaling was applied to each spectra individually. A binned, lower 
resolution spectrum with higher S/N was extracted for each spectra using 
optimal weighting, excluding parts of spectra contaminated by strong sky 
emission or strong atmospheric absorption. The resulting spectral resolutions 
of the binned X-shooter spectra were $R\approx 30-45$, $15-40$, and $20-50$ for 
the UVB, VIS, and NIR arms, respectively, while the spectral resolutions of 
the binned NIRSPEC spectra were $R\approx 30-100$ and $100-300$ for the $H$ 
and $K$ bands, respectively. 

\subsection{ACS-G800L Grism Spectrum}

C1-23152 lies within the area observed with the {\it Hubble Space Telescope} 
(HST) ACS G800L grism by the 3D-HST survey \citep{brammer12}; unfortunately it 
lies outside of the area covered by the infrared G141 grism. The G800L grism 
covers the wavelength range 0.55-1.0~$\mu\mathrm{m}$ with a dispersion of 
40\,\AA~pixel$^{-1}$ and a resolution of 80\,\AA\ for point-like sources 
\citep{kummel11}. We extract a flux-calibrated spectrum for C1-23152 using the 
\texttt{aXe} software \citep{kummel09}. The G800L spectrum of C1-23152 has no 
significant contamination from overlapping spectra of nearby sources and has a 
total integration time of 3123~s.

\subsection{ACS I$_{\rm 814}$ and WFC3 H$_{\rm 160}$ Imaging}
C1-23152 has been observed with HST using the WFC3 and the F160W filter 
(H$_{\rm 160}$, hereafter) as part of the HST Cycle 20 program GO-12990 (PI: 
Muzzin). Four individual H$_{\rm 160}$ exposures were aligned and combined with 
the \textit{MultiDrizzle} software \citep{koekemoer02} following the procedure 
outlined by \citet{skelton14}. The combined H$_{\rm 160}$ image has an exposure 
time of 861~s and is drizzled to a pixel scale of 
$0\farcs03~\mathrm{pixel}^{-1}$ with a point-source 
$\mathrm{FWHM}\sim0\farcs15$. ACS F814W (I$_{\rm 814}$, hereafter) 
imaging of C1-23152 is also available from the HST Cycle 12 program GO-9822 
\citep{scoville07}.


\section{ANALYSIS} \label{sec-analysis}

\subsection{Emission Features and Spectroscopic Redshift}

The nebular emission lines that we set out to measure were Ly{$\alpha$}, 
CIV{$\lambda$}1549, HeII{$\lambda$}1640, [OII]{$\lambda$}3727, H{$\beta$} 
and  [OIII]{$\lambda\lambda$}4959,5007. Upon visual inspection of the reduced 
two-dimensional spectra, the Ly{$\alpha$} (both X-shooter and Osiris) and 
[OIII]{$\lambda\lambda$}4959,5007 lines were easily identified. 

The emission lines (except Ly$\alpha$) were fit by a symmetric Gaussian 
profile plus a local continuum. In the case of the [OIII] doublet (easily 
identifiable by eye), identical redshift and FWHM were assumed with the ratio 
of the amplitudes [OIII]{$\lambda$}4959/[OIII]{$\lambda$}5007 fixed to 1:3.

The [OII]$\lambda\lambda$3726,3729 line doublet in the $H$-band was fit as a 
single Gaussian emission line, based on signal-to-noise and spectral resolution 
considerations. The analysis of [OII]{$\lambda$}3727 and H$\beta$ lines was 
more complicated due to significant sky residuals at their locations. 
Therefore, their measurements were carried out in multiple approaches. First, 
we fit both lines with redshift and FWHM (velocity) fixed at the best-fit 
values obtained from [OIII]. We then left the width as a free parameter. 
The [OII] line is only marginally detected in 
emission and we obtained a 3{$\sigma$} upper limit for the H$\beta$ emission 
by assuming the amplitude of the emission line to be three times the 
uncertainty in the local continuum.  CIV is detected in emission blue-shifted 
with respect to the redshift derived from [OIII] (this is discussed later in 
this section). HeII is marginally detected with a possible signature of 
self-absorption (also seen for Ly$\alpha$; see below). We determined the 
redshift of the self-absorption component using the central wavelength of the 
best-fit Gaussian line profile in absorption on top of an emission component. 
 
A Monte Carlo approach was used to measure the uncertainties in the centroid, 
flux, and width. For each spectra, 1000 simulated spectra were created by 
perturbing the flux at each wavelength of the true spectrum by a Gaussian 
random amount with the standard deviation set by the level of the 1{$\sigma$} 
error spectrum. Line measurements were obtained from the simulated spectra 
in the same manner as the actual data. We compute the formal lower and upper 
confidence limits by integrating the probability distribution of each 
parameter (centroid, width, continuum, and emission line flux) from the 
extremes until the integrated probability is equal to  0.1585.

The best fit values of the modeled spectral lines are listed in 
Table~\ref{tab-spec}, with the quoted uncertainties corresponding to the 
1$\sigma$ errors estimated from the Monte Carlo simulations.

\begin{deluxetable*}{lcccccccccc}
\centering
\tablecaption{Spectral line properties \label{tab-spec}}
\tablehead{\colhead{Feature}   & \colhead{$\lambda_{\rm lab}$}   & 
           \colhead{$\lambda_{\rm obs}$}   &  \colhead{$\sigma_{\rm obs}$}   & \colhead{$\sigma_{\rm inst}$} & \colhead{$\sigma_{\rm int}$} &
           \colhead{$z$} & \colhead{FWHM} & \colhead{$EW_{\rm obs}$} & \colhead{$L$} & \colhead{$\chi^{2}_{\rm red}$} \\
                     &  (\AA) &  (\AA) &  (\AA) & (\AA) & (\AA) &  & (km~s$^{-1}$) & (\AA) & ($10^{42}$~ergs~s$^{-1}$) & }
\startdata
\multicolumn{10}{l}{\it Keck-NIRSPEC} \\
OIII    & 4960.295 & 21578.9$^{+3.91}_{-7.67}$ & 25.1$^{+6.66}_{-6.60}$  & 6.02 & 24.4$^{+6.66}_{-6.60}$ & 3.351$^{+0.001}_{-0.002}$ & 798.0$^{+218.0}_{-215.9}$  & 37.9$^{+7.3}_{-6.6}$  &5.5$^{+5.0}_{-2.9}$ & 0.31 \\
OIII    & 5008.240 & 21792.0$^{+3.91}_{-7.67}$ & 25.1$^{+6.66}_{-6.60}$  & 6.02 & 24.4$^{+6.66}_{-6.60}$ & 3.351$^{+0.001}_{-0.002}$ & 790.2$^{+215.9}_{-214.90}$  & 113.8$^{+21.9}_{-19.8}$ &16.5$^{+15.1}_{-8.8}$ & 0.31 \\
OII$^{a}$   & 3727.80  & 16220.5 & 19.13 & 6.24 & 18.08 & 3.351 & 786.9 & 54.4$^{+29.2}_{-23.4}$  & 7.54$^{+10.1}_{-4.17}$  & 0.18 \\

H$\beta^{a}$ & 4862.68  & 21158.7 & 24.4 & 6.24 & 23.6 & 3.351 & 786.9 & -14.7$^{+13.9}_{-14.6}$ &  $\le\;0.88$ & 0.30 \\
\\
\multicolumn{10}{l}{\it VLT-X-shooter} \\
Ly$\alpha$  & 1215.24  & 5290.47$\pm 0.57 $ & 6.93$\pm 0.6$ & 0.45 & 6.92$\pm 0.6$ & 3.353$\pm 0.001$ & 923.0$\pm$79.3 & 466.4$\pm$ 231.3 & 46.2$\pm$6.8 & 0.47\\
Ly$\alpha_{self-abs}$ & 1215.24 & 5290.58$^{+1.53}_{-0.73}$ &\nodata &\nodata &\nodata & 3.354$^{+0.001}_{-0.001}$  &\nodata &\nodata &\nodata  & 0.53\\
Ly$\alpha$1 &\nodata   & 5284.6$^{+0.70}_{-0.57}$ & 1.75$^{+0.76}_{-1.15}$ & 0.45 & 1.69$^{+0.76}_{-1.15}$ & 3.349$^{+0.001}_{-0.001}$ &\nodata  & \nodata & \nodata & 0.50\\
Ly$\alpha$2 & \nodata  & 5293.6$^{+1.40}_{-1.71}$ & 8.01$^{+2.46}_{-2.28}$ & 0.45 & 8.00$^{+2.46}_{-2.28}$ & 3.356$^{+0.001}_{-0.002}$ &\nodata  & \nodata & \nodata & 0.50\\
HeII       & 1640.405 & 7132.4$^{+4.09}_{-3.64}$  & 5.49$^{+2.52}_{-3.27}$ & 0.34 & 5.48$^{+2.52}_{-3.27}$ & 3.348$^{+0.003}_{-0.002}$   &542.1$^{+249.5}_{-323.4}$  & 48.3$^{+22.8}_{-26.8}$ & 8.46$^{+10.6}_{-8.0}$ & 0.36 \\
HeII$_{self-abs}$ & 1640.405 & 7132.5$^{+4.09}_{-3.64}$ & \nodata & \nodata & \nodata & 3.348$^{+0.001}_{-0.001}$ & \nodata & \nodata & \nodata & 0.35\\
OIII        & 4960.295 & 21570.0$^{+15.5}_{-8.10}$  & 18.16$^{+7.32}_{-9.82}$  & 1.74 & 18.08$^{+7.32}_{-9.82}$ & 3.349$^{+0.003}_{-0.002}$   & 591.6$^{+239.4}_{-239.4}$  & 80.1$^{+17.7}_{-36.4}$ & 15.0$^{+5.8}_{-1.4}$ & 0.49\\  
OIII        & 5008.240 & 21778.5$^{+15.5}_{-8.10}$ & 18.16$^{+7.32}_{-9.82}$  & 1.74 & 18.08$^{+7.32}_{-9.82}$ & 3.349$^{+0.003}_{-0.002}$ & 585.9$^{+237.3}_{-318.3}$  & 240.2$^{+53.1}_{-109.0}$ & 45.0$^{+17.0}_{-4.3}$  & 0.49\\ 
\\
\multicolumn{10}{l}{\it GTC-Osiris} \\
Ly$\alpha$   & 1215.24 & 5289.14$\pm 0.31 $  & 5.32$\pm 0.32$     & 3.47 & 4.03$\pm 0.32$ &  3.352$\pm 0.001$ & 538.4$\pm$42.1 & 485.0$\pm$145.0 & 34.3$\pm$2.7 & 1.53\\
Ly$\alpha_{self-abs}$ & 1215.24 & 5288.3$^{+0.6}_{-0.7}$ & \nodata   & \nodata   & \nodata   & 3.352$^{+0.001}_{-0.001}$  &\nodata &\nodata &\nodata  & 1.89\\
Ly$\alpha$1  & \nodata  & 5283.1$^{+0.5}_{-0.6}$  & 2.57$^{+0.99}_{-0.84}$ & 3.47 &\nodata & 3.347$^{+0.001}_{-0.001}$ &\nodata  & \nodata & \nodata & 1.63\\
Ly$\alpha$2  & \nodata  & 5293.9$^{+0.6}_{-0.6}$  & 2.94$^{+0.68}_{-0.53}$ & 3.47 &\nodata & 3.356$^{+0.001}_{-0.001}$ &\nodata  & \nodata & \nodata & 1.63\\
CIV         & 1549.48   & 6732.6$^{+2.3}_{-1.9}$  & 7.44$^{+2.04}_{-1.85}$ & 4.35 & 6.04$^{+2.04}_{-1.85}$ & 3.345$^{+0.002}_{-0.001}$   &633.6$^{+214.0}_{-194.1}$  & 40.8$^{+21.7}_{-10.3}$  &  3.35$^{+1.17}_{-1.71}$ & 1.44  \\
CIV$_{\rm abs}$ & 1549.48 & 6706.1$\pm$1.9 & 6.06$^{+2.37}_{-4.14}$ & 4.35 & 4.23$^{+2.37}_{-4.14}$ & 3.328$\pm$0.001 & 445.4$^{+249.6}_{-436.0}$ & -24.0$^{+16.0}_{-8.4}$ & \nodata & 1.44\\
\enddata   
\tablecomments{$\lambda_{lab}$ is the laboratory rest-frame wavelength of 
the targeted spectral lines; $\lambda_{obs}$ is the observed wavelength of 
the corresponding lines; $\sigma_{obs}$ is the observed width of the modeled 
spectral lines; $\sigma_{inst}$ is the instrumental width determined from the 
sky lines; $\sigma_{int}$ is the intrinsic width of the spectral lines in the 
observed frame; $z$ is the derived redshift; FWHM is the intrinsic velocity 
width of the spectral lines; $EW_{\rm obs}$ is the observed equivalent widths 
of the spectral lines; and $L$ is the integrated line luminosity calculated 
using the adopted systemic redshift $z_{\rm spec}=3.351$. $\chi^{2}_{\rm red}$ is the 
reduced $\chi^{2}$ values for corresponding emission line fits. Listed errors 
correspond to the 1$\sigma$ uncertainties derived from the Monte Carlo 
simulations. $^{a}$ denotes fits to emission lines with redshift {\it and} 
line width fixed to that of [OIII]. }
\end{deluxetable*}

The profile of the Ly{$\alpha$} emission is observed to be asymmetrically 
double-peaked in both the X-shooter and Osiris spectra, indicative of 
self-absorption, and we analyzed it with a number of approaches. We initially 
fit each peak independently with a Gaussian profile, assuming a constant 
continuum. Then, in order to account for the self-absorption, we fit the 
observed spectra simultaneously with two Gaussians, one in emission and the 
other in absorption. Finally, we fit the wings of the Ly{$\alpha$} emission 
with a single Gaussian after masking the self-absorbed region between the two 
peaks in order to obtain the line width of the emission line. We determine 
that the higher S/N X-shooter spectrum samples the wings of the Ly{$\alpha$}
emission line better, resulting in a FWHM of $\sim923.0\pm79.3$ km~s$^{-1}$, 
comparable with the widths found in other high-$z$ Type II QSOs 
\citep{norman02,alexandroff13}. 

\begin{figure*}
\epsscale{1}
\plotone{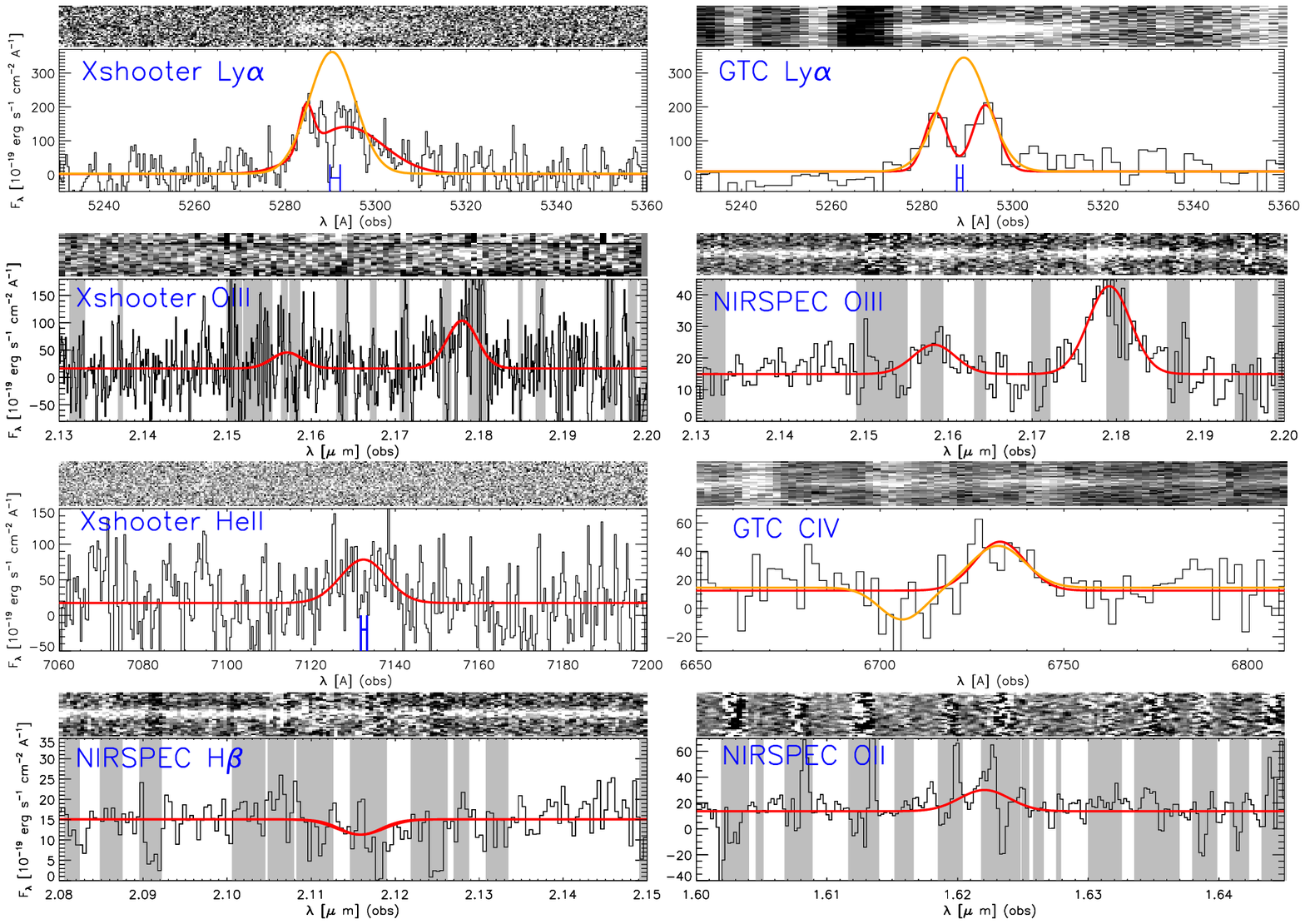}
\caption{Observed one-dimensional and two-dimensional spectra in the region around the considered 
spectral features. The red solid curves represent the best-fit Gaussian 
profiles to the emission lines in 1D spectra. The orange solid curves in the top panels 
represent the best-fit single Gaussian profile after masking the self-absorbed 
region between the two Ly$\alpha$ peaks. The orange solid curve in the CIV 
panel represents best-fit P-Cygni profile. The gray shaded regions indicate 
regions of the spectra significantly affected by telluric sky lines. The best 
fit central wavelength and 1$\sigma$ uncertainty of the central wavelength of 
the self-absorption component of Ly$\alpha$ and HeII is indicated by the blue 
brackets in the corresponding panels. 
\label{fig-fits_plots}}
\end{figure*}

\begin{figure}
\epsscale{1.15}
\plotone{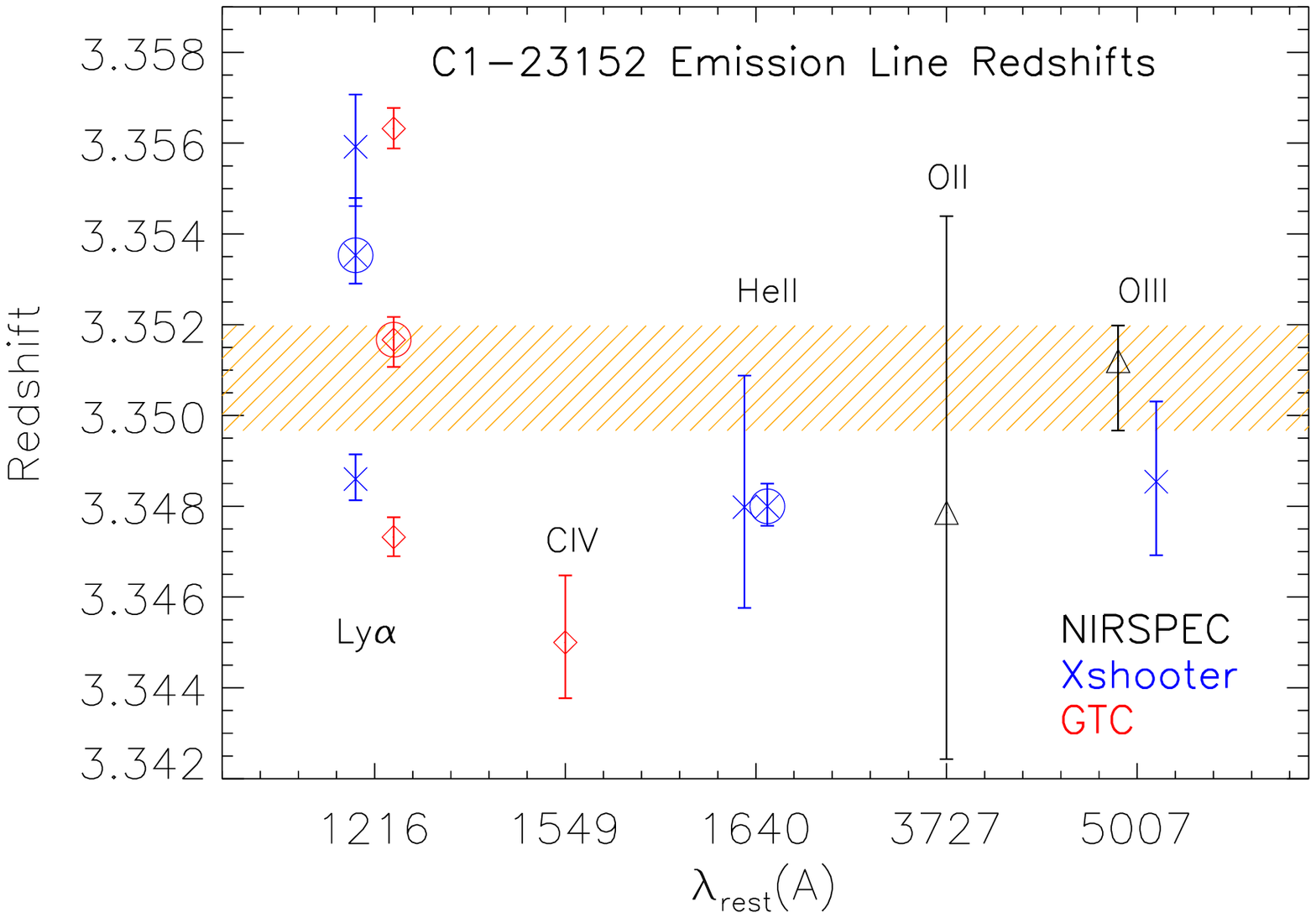}
\caption{Spectroscopic redshifts obtained from different emission lines. 
\emph{Blue crosses} show the redshifts of spectral features measured in the 
X-shooter spectra; \emph{red diamonds} show the redshifts of spectral features 
measured in the Osiris spectra; \emph{black triangles} indicate the redshifts 
of spectral features measured in the NIRSPEC spectra. Open circles represent 
the redshifts measured from the self-absorption features in the Ly$\alpha$ and 
HeII lines. The orange hatched region indicates the adopted systemic 
redshift for C1-23152 based on the higher S/N NIRSPEC [OIII] detection, i.e., 
$z_{\rm spec}=3.351^{+0.001}_{-0.002}$.\label{fig-Z_plots}}
\end{figure}

In Figure~\ref{fig-fits_plots} we show the observed one-dimensional spectra 
around the regions of the considered spectral lines. The regions of the spectra 
significantly affected by sky lines (shown by gray filled areas) were excluded 
from the line profile modeling. The redshift of a galaxy's nebular lines 
(e.g., [OII]$\lambda$3727, H$\beta$, [OIII]$\lambda \lambda$4959,5007) is 
expected to be nearly equal to the redshift of its stars, i.e., the gas 
responsible for the nebular emission should always lie close to hot stars. We 
therefore choose the redshift of the [OIII] emission lines, 
$z_{\rm spec}=3.351^{+0.001}_{-0.002}$, to be the systemic redshift of the galaxy.

Due to the resonant nature of the transition in neutral hydrogen, the 
interpretation of the origin of the Ly$\alpha$ emission is non-trivial 
\citep{verhamme06, kulas12}. In Figure~\ref{fig-Z_plots} we show the 
spectroscopic redshifts and corresponding uncertainties for all observed 
spectral lines. We note that the prominent double peaked Ly$\alpha$ in both 
the X-shooter and Osiris spectra brackets the systemic redshift of the galaxy, 
and the redshifts of the self-absorbing material obtained from both the 
Ly$\alpha$ and HeII are also consistent with the systemic redshift derived from 
[OIII].

We marginally detect an emission line with a potential P-Cygni profile at 
6732\AA~in the Osiris spectrum, which we interpret as CIV$\lambda$1549 
emission blueshifted with respect to the systemic redshift by 425~km~s$^{-1}$. 
P-Cygni line profiles are powerful diagnostics of outflow kinematics 
\citep{spoon13}. We fit the observed P-Cygni profile of CIV with a component in 
emission and a blue-shifted broad component in absorption. The absorption 
feature of the P-Cygni profile is blue-shifted by $\sim$1160$\pm$130 
km~s$^{-1}$ with respect to the emission component. We note that 
high-ionization emission lines blue-shifted with respect to low-ionization 
emission lines have been previously observed in (particularly radio-quiet) 
luminous quasars \citep{richards11}, who argue that the apparent 
\emph{blueshift} with respect to the systemic redshift might be the result of 
preferential reduction/obscuration of the red wing of the CIV emission line. 
We however stress that the CIV detection in the Osiris spectrum is only marginally 
significant, and that CIV emission is not detected in the X-shooter spectrum, 
therefore caution is required when interpreting the ambiguous CIV detection. 
Although a CIV emission feature is often used as evidence for the 
identification of active galactic nuclei (AGNs; \citealt{daddi04}, 
\citealt{hainline11}), it has been found to be missing in $z \sim 2.5$ AGNs 
with moderately broad (FWHM $\sim$ 1430~km~s$^{-1}$) Ly$\alpha$ emission 
\citep{hall04}. 

We calculated the fluxes of the observed emission lines by integrating the 
fitted Gaussian profiles, with uncertainties determined using the 1$\sigma$ 
standard deviations derived from Monte Carlo simulations of the amplitude and 
velocity width of the fits. The line luminosities estimated using the adopted 
systemic redshift are listed in Table~\ref{tab-spec}, which also lists the 
line velocities (FWHM) corrected for the instrumental profile (determined 
from the width of the sky lines). The [OIII] line width is 
FWHM$_{\rm{[OIII]}}\approx$790$\pm$220~km~s$^{-1}$, in good agreement with the 
line widths observed in low-redshift Seyfert II galaxies 
(FWHM$<$1200~km~s$^{-1}$; \citealt{hao05}). The Ly$\alpha$ line width is 
slightly broader, with FWHM$_{\rm{Ly\alpha}}\approx$920$\pm$80~km~s$^{-1}$.


\subsection{SED Modeling and Stellar Population Properties} \label{sec-sedmodel}

In this section we present the modeling of the observed SED of C1-23152 
obtained by combining the photometry from NMBS and the spectra from the UV to 
the NIR, and we present the derived properties of its stellar population. 

In order to robustly constrain the stellar population parameters, all spectra 
and photometry must be corrected for contamination from nebular emission 
lines. We determined the necessary emission-line corrections for all observed 
photometries by comparing the observed-frame equivalent width of each emission 
line to the bandwidth of corresponding filter. We found that the contributions 
due to [OII] and [OIII] amount to $< 5\%$  in the NMBS $H_{1}$, $H$, $K$, and 
$K_{\rm S}$ bands. The contributions due to Ly$\alpha$ emission were much more 
significant in the \textit{g} and \textit{V} broad-bands, $\sim$60\% and 
$\sim$90\%, respectively. We also removed the photometry belonging to the 
medium-band IA527 (Subaru) from the fitting routine as this band was 
completely dominated by the Ly$\alpha$ emission. 

\begin{figure*}
\epsscale{1}
\plotone{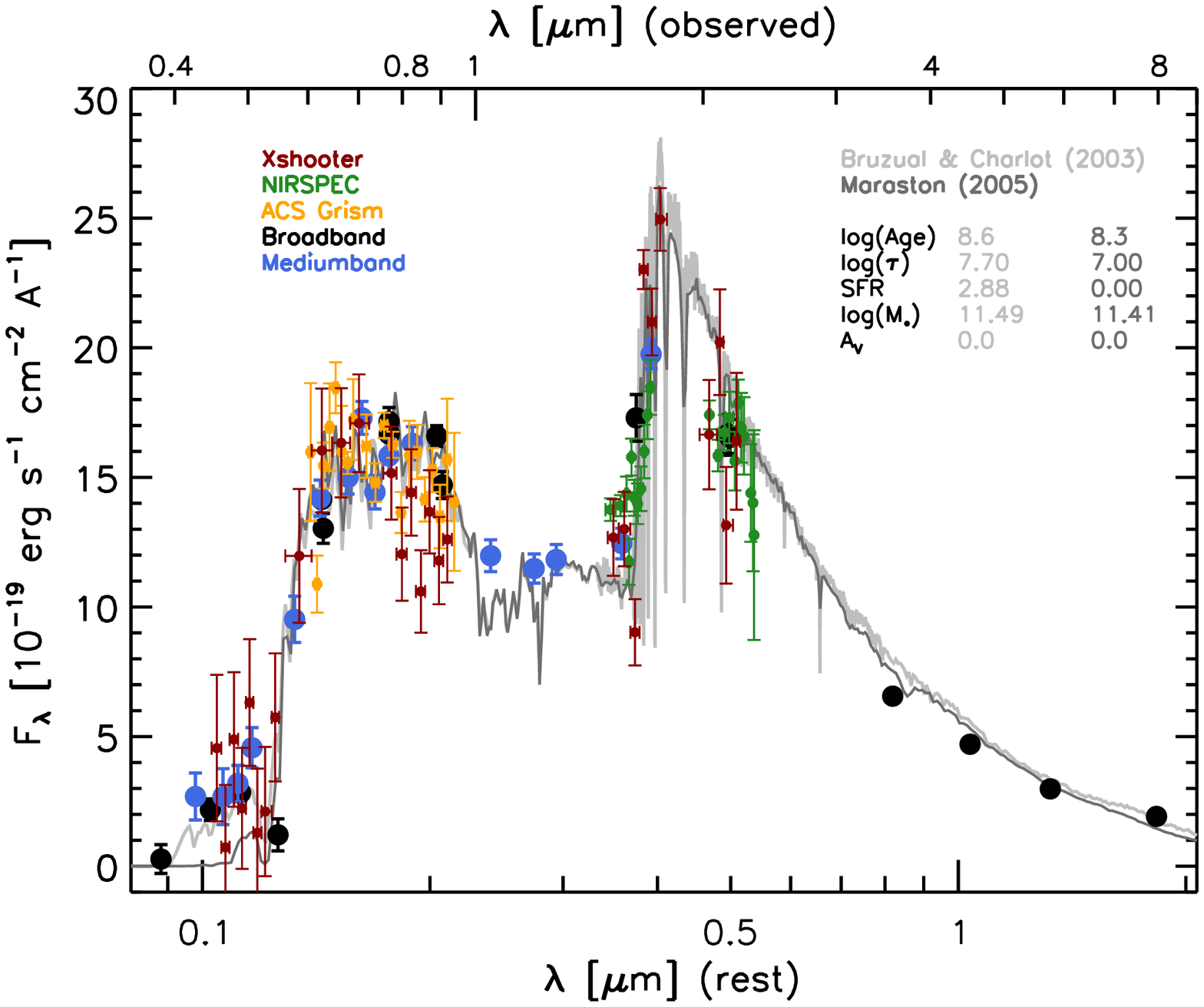}
\caption{Observed SED from the combination of the medium- (blue filled 
circles) and broad-band (black filled circles) NMBS photometry and the binned 
spectra (NIRSPEC in green; ACS grism in orange; and X-shooter in maroon) 
redshifted to the rest frame. The light and dark gray curves represent the 
best-fit models adopting the BC03 and MA05 stellar population models, 
respectively, with an exponentially declining SFH and free metallicity. 
Best-fit stellar population properties of the BC03 and MA05 models are indicated
in legend. 
\label{fig-FAST_originalwall}}
\end{figure*}

We estimated the stellar population properties by fitting the binned, 
low-resolution UVB, visual, and NIR X-shooter, NIRSPEC, and ACS grism spectra 
in combination with the broad-band and medium-band photometry with stellar 
population synthesis (SPS) models. We excluded the Osiris spectrum due to its 
lower S/N compared to the binned X-shooter spectrum and the photometry in the 
same wavelength regime. We used FAST (Fitting and Assessment of Synthetic 
Templates; \citealt{kriek09}) to model and fit a full grid in metallicity, 
dust content, age, and star formation timescale. We investigated different 
star formation histories (SFH), i.e., the exponentially declining, the 
delayed-exponentially declining, and the truncated SFHs. We adopted stellar 
population synthesis models from both \citet{bc03} and \citet{maraston05} 
(BC03 and MA05, hereafter), and we assumed the \citet{kroupa01} IMF and the 
\citet{calzetti00} extinction law. The age range allowed was between 10~Myr 
and 1.9~Gyr (the maximum age of the universe at the redshift of the galaxy) 
with a step size of 0.1~dex. We adopted a grid for $\tau$ between 3.16~Myr and 
10~Gyr in steps of 0.10 dex. We allowed the dust attenuation ($A_{\rm V}$) to 
range from 0 to 3 mag with step size of 0.01~mag. Metallicity can vary between 
$Z=0.004$, 0.08, 0.02, 0.05 or $Z=0.001$, 0.01, 0.02, 0.04 for the BC03 and 
the MA05 templates, respectively. We initially modeled the observed SED with 
the metallicity as a free parameter and then we repeated the modeling by 
treating the metallicity as a systematic uncertainty by fixing it at the 
available values.

Figure~\ref{fig-FAST_originalwall} shows the observed SED blueshifted to the 
rest-frame of the galaxy with the MA05 and BC03 stellar population synthesis 
models and an exponentially declining SFH. The results of the SED-modeling and 
the corresponding 3$\sigma$ errors are listed in Table~\ref{tab-sed}. 

As shown by the values in Table~\ref{tab-sed}, the median stellar mass derived 
adopting the BC03 models is $\log{(M_{*}/M_{\sun})}=11.49\pm0.08$. The error 
budget is dominated by the systematic uncertainties due to different 
SED-modeling assumptions. Nonetheless, given the very high-quality and 
sampling of the SED of C1-23152, different modeling assumptions affect the 
derived stellar mass by at most $\pm$0.08~dex, i.e., $<$20\%. Independently of 
the SED-modeling assumptions, the best-fit dust attenuation is zero, with a 
3$\sigma$ upper limit of $\sim$0.4~mag when the observed SED is modeled with 
MA05 stellar population synthesis models, a delayed-exponentially declining 
SFH, and super-solar metallicity. This combination of SED-modeling assumptions 
also results in the largest 3$\sigma$ upper limit for the SFR, i.e., 
7~M$_{\sun}$~yr$^{-1}$. However, most of the other combinations of SED-modeling 
assumptions result in low or negligible SFRs, with a more typical 3$\sigma$ 
upper limit of 2-5~M$_{\sun}$~yr$^{-1}$. The best-fit SFR-weighted mean stellar 
age $\langle t \rangle_{SFR}$ of the galaxy (as defined in 
\citealt{forsterschreiber04}) is 350~Myr when adopting BC03 models and a 
best-fit solar metallicity, and a factor of 2 smaller when adopting MA05 
models with a best-fit super-solar metallicity. We 
note that the stellar age 
can vary by as much as $\pm$0.2~dex depending on the metallicity. The 
preferred timescale of the duration of the burst ($\tau$) is always very 
short, $<$70~Myr at 3$\sigma$, with a typical best-fit value of $\sim$50~Myr. 
We finally note that the value of $\tau$ is found to be systematically lower 
by as much as a factor of $\sim$8 when the MA05 models are assumed. The 
best-fit stellar age of $\sim$400~Myr implies a formation redshift of 
$z_{\rm form}\sim4.1$.

It should be noted that the stellar population parameters derived above likely 
suffer from additional systematic errors due to uncertainties in the modeling of the evolution of 
stellar populations (e.g., differences in stellar population synthesis models; see, e.g., \citealt{marchesini09, conroy09, conroy10a, conroy10b}) 
and variations in the assumed IMF. \citet{conroy09} find that stellar 
mass estimates of $z\sim2$ luminous red galaxies can vary by $\sim$ 0.6 dex 
due to different models 
of the advanced stages of stellar evolution, particularly the thermally pulsating 
asymptotic branch (TP-AGB) phase. Assuming a different IMF will also affect the calculated stellar mass 
as this affects the stellar mass-to-light ratio ($M/L$), with it being more 
sensitive to age for more top-heavy IMF. For example, using a relatively more bottom heavy \citet{salpeter55} 
IMF than the adopted \citet{kroupa01} would result in stellar masses $\sim$ 0.2 dex higher. Recently, several studies 
(e.g., \citealt{conroy12, conroy13, spiniello13} concluded that the IMF becomes increasingly 
bottom heavy toward galaxies characterized by larger velocity dispersion. Given the inferred 
large stellar mass and its compact size ($\sim~1$ kpc; see Section~\ref{sec-size}), it is plausible 
that a super-Salpeter IMF may be more appropriate for C1-23152, resulting in an even larger
stellar mass (by $\sim$0.3 dex with respect to Kroupa; \citealt{conroy12}). 

Compared to the stellar population parameters derived in \citet{marchesini10} 
for C1-23152 based only on the NMBS photometry, we find that the stellar mass 
is larger by $\sim 0.07$ dex, SFR larger by $\sim 1.5$ dex (but well within 
the quoted uncertainties listed in \citealt{marchesini10}) and the stellar age 
older by $\sim 0.5$ dex.

Given the evidence for more complex SFHs in high-$z$ galaxies found in literature 
(e.g., \citealt{papovich11, finkelstein11}) we also consider a scenario in which a recent burst 
of star formation is triggered through gas infall. To investigate this, we fit the 
observed SED with libraries created using GALAXEV \citep{bc03} for two-component 
SFHs. We use exponentially declining SFHs (with log($\tau$ [yr])= 7.1, 7.3, 7.5, 7.7, 7.9) 
to model the old stellar population accounting for the majority of the galaxy's stellar mass, 
and introduce a burst contributing a fraction of the stellar mass ($5\%$, $10\%$, $20\%$, 
$30\%$ or $50\%$) when the old stellar population reaches an age of 0.2, 0.3, 0.7, 1.0, 
1.2 or 1.4 Gyrs. Each SFH is modeled for four different metallicities 
(Z=0.004, 0.008, 0.02, 0.05). We used FAST to fit the observed SED using these libraries 
in two manners. First, we modeled the observed SED with only an upper limit restriction 
to the age of the stellar population dictated by the age of the universe at $z=3.3512$. In this case, 
the modeled fits for the SFHs with $\Delta t_{\rm burst}$~$\geq$~0.3~Gyr are essentially identical, with 
$\log(M_{*}/M_{\odot})$=~11.42, SFR=4$M_{\odot}$yr$^{-1}$ and an age of $\sim250$~Myr, 
without bursts taking place. The only SFHs in which the fits favor the occurence of the burst.  
are the $\Delta t_{\rm burst}=0.2$~Gyr models. The median stellar mass is
$\log(M_{*}/M_{\odot})$=~11.48$\pm0.2$, and SFR=1.5$\pm$1.5~$M_{\odot}$yr$^{-1}$. The age 
of the stellar population for these SFHs is $\sim400$~Myr, indicating that the bursts 
occur $\sim200$~Myr prior to the time of observation. For the second run of FAST 
we additionally restricted the lower limit of the ages of the libraries used to fit the 
observed SED, such that a burst is implicitly forced to occur in these cases. 
We found that $\Delta t_{\rm burst}$~$\geq$~0.71~Gyr models 
are not able to reproduce the observed SED well, hence we reject 
these scenarios as plausible SFHs for C1-23152. For the $\Delta t_{\rm burst}=0.2$~Gyr 
and $\Delta t_{\rm burst}=0.3$~Gyr models the best-fit stellar masses are $\log(M_{*}/M_{\odot})$=~11.46$\pm0.03$
and $\log(M_{*}/M_{\odot})$=~11.51$\pm0.03$, respectively. For both SFHs, the best-fit
ages indicate that the bursts take place $\sim150-200$~Myr prior to the epoch of
observation, setting the formation redshifts to be $z_{\rm form}\sim$~3.9-4.1 for
$\Delta t_{\rm burst}=0.2$~Gyr and $z_{\rm form}\sim$~4.1-4.5 for $\Delta t_{\rm burst}=0.3$~Gyr SFHs. 
Although the SED fits for the two-component SFHs systematically return higher $\chi^{2}$ values 
compared to more simple exponentially declining or delayed SFHs, the best-fit
stellar population parameters for $\Delta t_{\rm burst}$~$\leq$~0.3~Gyr SFHs are consistent.

It is worth noting that the observed SED of C1-23152 bears a striking resemblance to 
the spectrum of Vega, an A0V star. It is plausible to assert the SFH of C1-23152 is 
not composite as its age coincides with the time necessary for the turn-off point 
of main sequence stars for a single stellar population to have almost reached AV0 stars.  
In this scenario the observed SED would be dominated by light from AV0 stars, 
explaining the similarity between the observed SED and spectrum of Vega. This serves as an 
additional constraint on the age and SFH of C1-23152.

\begin{deluxetable*}{lcccccccc}
\centering
\tablecaption{Best Fit Stellar Population Parameters\label{tab-sed}}
\tablehead{   & \colhead{$\log{(\tau)}$} & \colhead{Metallicity}   & \colhead{$\log{(Age)}$}   & \colhead{$A_{\rm V}$} &
           \colhead{$\log{(M_{*})}$}   &  \colhead{SFR}   & \colhead{sSFR} & \colhead{$\chi^2$} \\
                 & (yr)  &   & (yr)  & (mag) & (M$_{\sun}$) & (M$_{\sun}$~yr$^{-1}$)  & log($yr^{-1}$) &  }
       
\startdata
\multicolumn{9}{l}{\it SFH: Exponentially declining} \\
Free Z & $7.70_{-0.30}^{+0.02}$ &  $ 0.02_{-0.003}^{+0.030}$ & $8.60_{-0.23}^{+0.02}$ & $0.00_{-0.00}^{+0.14}$ & $11.49_{-0.09}^{+0.00}$ & $2.88_{-2.28}^{+1.01}$ & $ -11.03_{-0.58}^{+0.20}$ & $3.56$ \\

Free Z  & $(7.00_{-0.23}^{+0.40})$   &  $(0.04_{-0.006}^{+0.000})$  & $(8.30_{-0.06}^{+0.03})$  & $ (0.0_{-0.0}^{+0.2})$  & $(11.41_{-0.03}^{+0.05})$   & $(0.00_{-0.00}^{+5.13}) $ & $(-15.40_{-4.54}^{+4.66}) $ & $(3.89)$  \\
           Z=0.004    & $7.80_{-1.30}^{+0.03}$   &  $0.004$  & $8.80_{-0.10}^{+0.01}$  & $0.00_{-0.00}^{+0.05}$  & $11.56_{-0.05}^{+0.00}$   & $ 0.37_{-0.37}^{+0.00} $ & $ -11.99_{-87.01}^{+0.00} $ & $3.86$ \\

           Z=0.008   & $7.70_{-1.20}^{+0.10}$   &  $0.008$  & $8.70_{-0.12}^{+0.01}$  & $0.00_{-0.00}^{+0.12}$  & $11.54_{-0.07}^{+0.00}$   & $ 0.44_{-0.44}^{+1.91} $ & $ -11.90_{-87.10}^{+0.78} $ & $3.80$ \\

           Z=0.02    & $7.70_{-0.64}^{+0.02}$   &  $0.02$  & $8.60_{-0.10}^{+0.01}$  & $0.00_{-0.00}^{+0.09}$  & $11.49_{-0.06}^{+0.00}$   & $ 2.88_{-2.88}^{+0.00} $ & $ -11.03_{-6.57}^{+0.00} $ & $3.52$ \\

           Z=0.05    & $7.50_{-0.10}^{+0.00}$   &  $0.05$  & $8.40_{-0.03}^{+0.00}$  & $0.10_{-0.10}^{+0.04}$  & $11.42_{-0.02}^{+0.00}$   & $ 3.89_{-3.29}^{+0.00} $ & $ -10.83_{-0.78}^{+0.00} $ & $3.52$ \\
\\
\hline
\\
\multicolumn{9}{l}{\it SFH: Delayed exponentially declining} \\
Free Z & $7.60_{-1.10}^{+0.15}$ &  $0.008_{-0.004}^{+0.042}$ & $8.70_{-0.33}^{+0.14}$ & $0.00_{-0.00}^{+0.13}$ & $11.49_{-0.08}^{+0.06}$ & $0.48_{-0.48}^{+4.65}$ & $-11.81_{-87.19}^{+1.09}$ & $3.63$ \\
Free Z & $(6.70_{-0.20}^{+0.70}) $   &  $(0.04_{-0.015}^{+0.000})$  & $(8.30_{-0.14}^{+0.10})$  & $(0.00_{-0.00}^{+0.42})$  & $(11.41_{-0.02}^{+0.11})$   & $(0.00_{-0.00}^{+6.76} )$ & $(-21.55_{-8.78}^{+10.92}) $ & $(3.92)$ \\

           Z=0.004    & $7.60_{-0.05}^{+0.10}$   &  $0.004$  & $8.80_{-0.05}^{+0.02}$  & $0.00_{-0.00}^{+0.03}$  & $11.55_{-0.03}^{+0.00}$   & $ 0.03_{-0.00}^{+0.4} $ & $ -13.08_{-0.00}^{+1.17} $ & $3.63$ \\

           Z=0.008   & $7.60_{-1.10}^{+0.20}$   &  $0.008$  & $8.70_{-0.13}^{+0.10}$  & $0.00_{-0.00}^{+0.16}$  & $11.49_{-0.03}^{+0.07}$   & $ 0.48_{-0.48}^{+3.07} $ & $ -11.81_{-87.19}^{+0.80} $ & $3.59$ \\

           Z=0.02    & $7.50_{-0.99}^{+0.10}$   &  $0.02$  & $8.60_{-0.17}^{+0.02}$  & $0.00_{-0.00}^{+0.18}$  & $11.48_{-0.06}^{+0.00}$   & $ 0.58_{-0.58}^{+4.10} $ & $ -11.72_{-25.02}^{+0.90} $ & $3.71$ \\

           Z=0.05    & $7.30_{-0.77}^{+0.20}$   &  $0.05$  & $8.40_{-0.14}^{+0.10}$  & $0.10_{-0.10}^{+0.22}$  & $11.41_{-0.03}^{+0.03}$   & $ 0.51_{-0.51}^{+3.75} $ & $ -11.52_{-13.89}^{+0.80} $ & $3.76$ \\
\\
\hline
\\
\multicolumn{9}{l}{\it SFH: Truncated} \\
Free Z & $8.20_{-1.70}^{+0.43}$ & $0.008_{-0.004}^{+0.042}$ & $8.70_{-0.46}^{+0.23}$ & $0.00_{-0.00}^{+0.33}$ & $11.50_{-0.12}^{+0.08}$ & $0$ &  $-99$ & $3.62$ \\
 & $(7.40_{-0.90}^{+0.93})$ &  $(0.04_{-0.039}^{+0.000})$  & $(8.30_{-1.30}^{+0.17})$  & $(0.00_{-0.00}^{+2.50})$  & $(11.41_{-0.04}^{+0.12})$   & $(0)$ &  $(-99) $& $(4.54)$ \\
\\
\hline
\\
\multicolumn{9}{l}{\it AGN continuum subtracted; SFH: Exponentially declining} \\
\multicolumn{9}{l}{\it $\alpha\sim$1.6} \\
Free Z & $7.70_{-1.20}^{+0.82}$ &  $ 0.02_{-0.013}^{+0.03}$ & $8.60_{-0.3}^{+0.14}$ & $0.00_{-0.00}^{+0.10}$ & $11.44_{-0.01}^{+0.14}$ & $2.57_{-2.57}^{+0.00}$ & $ -11.03_{-99}^{+0.00}$ & $4.43$ \\

\multicolumn{9}{l}{\it Mrk231 template} \\
Free Z & $7.70_{-1.20}^{+0.05}$ &  $ 0.02_{-0.009}^{+0.03}$ & $8.60_{-0.34}^{+0.08}$ & $0.00_{-0.00}^{+0.25}$ & $11.46_{-0.14}^{+0.00}$ & $2.69_{-0.00}^{+0.94}$ & $ -11.03_{-21.05}^{+0.20}$ & $4.52$ \\

\multicolumn{9}{l}{\it DR2 template} \\
Free Z & $7.50_{-1.00}^{+0.10}$ &  $ 0.02_{-0.005}^{+0.03}$ & $8.60_{-0.21}^{+0.04}$ & $0.00_{-0.00}^{+0.07}$ & $11.49_{-0.11}^{+0.01}$ & $0.05_{-0.05}^{+0.41}$ & $ -12.80_{-25.70}^{+0.99}$ & $8.03$ \\

\multicolumn{9}{l}{\it TQSO1 template} \\
Free Z & $6.90_{-0.40}^{+0.73}$ &  $ 0.05_{-0.038}^{+0.00}$ & $8.40_{-0.04}^{+0.27}$ & $0.00_{-0.00}^{+0.02}$ & $11.39_{-0.01}^{+0.11}$ & $0.00_{-0.00}^{+0.46}$ & $ -20.03_{-18.47}^{+8.22}$ & $7.91$ \\

\enddata
\tablecomments{Estimated stellar population parameters from the modeling of 
the binned UV-to-NIR spectra in combination with the broad- and 
medium-bandwidth photometry from NMBS. The values in parenthesis correspond to 
the best-fit stellar population parameters assuming the \citet{maraston05} 
stellar population synthesis models, while the values not in parenthesis 
correspond to the best-fit parameters derived assuming the \citet{bc03} 
stellar population synthesis models. Quoted errors are 3$\sigma$ confidence 
intervals output by FAST (see \citealt{kriek09} for a detailed description of 
the adopted method in FAST to estimate confidence intervals). A 
\citet{kroupa01} IMF and a \citet{calzetti00} extinction law is assumed in all 
cases. }
\end{deluxetable*}

\subsection{Star Formation Rate from [OII]}

We used the emission line [OII]$\lambda$3727 as an independent measure of the 
SFR and compared it to the values obtained from SED fitting. We emphasize that 
the [OII] detection is very marginal due to significant skyline contamination 
in the spectral region where the redshifted [OII] line falls.  We used the L[OII]-
SFR relation derived in \citet{kewley04} and obtained SFR([OII])=
$30.9^{+42.1}_{-18.8};\rm{M_{\sun}\;yr^{-1}}$ (1$\sigma$ 
error), larger by a factor of $\sim $15 (for an 
exponentially-declining SFH) and $\sim$65 (for a delayed 
exponentially-declining SFH) compared to the best-fit SFR values from the SED 
modeling. Studies find that radiation from LINER or Seyfert components in post-starburst
and red sequence appear to be a more prominent source of [OII] line luminosity compared to star formation
processes \citep{lemaux10}.
Assuming the [OIII] line is purely of AGN origin in our spectra, we 
adopt an empirical relation between [OII] and [OIII] to remove the component 
of the detected [OII] line that can be attributed to the presence of an AGN. 
\citet{silverman09} used the ratio [OII]/[OIII]=0.21 found in type 1 AGN in 
the SDSS sample with log $L_{\rm [OIII]} > 41.5$  to provide an estimate of the 
[OII] emission line luminosity free of any AGN contribution. We use this 
value to calculate an AGN-corrected upper limit to the SFR of the galaxy based 
on [OII] emission line luminosity and obtain  
SFR([OII$_{\rm AGN corrected}$])= $17^{+44}_{-17}\;\rm{M_{\sun}\;yr^{-1}}$, 
consistent within the errors with the SFR derived from SED 
modeling. We note that while [OII]-derived SFR is very uncertain due to 
the large uncertainties in the [OII] line flux measurements, it is valuable as an 
independent tracer of the star formation rate.

\subsection{Emission Line Properties: AGN vs. Star Formation?}

One diagnostic of AGN in several studies has been the luminosity of the 
[OIII]$\lambda$5007 optical nebular emission. We confirm that C1-23152's 
[OIII] emission is indeed very luminous, $L_{\rm [OIII] }=1.65\times10^{43}$ 
erg~s$^{-1}$. This value is at the high 
end of the range of [OIII] luminosities found for Lyman Break galaxies (LBG) 
and Ly$\alpha$ emitters (LAE) at $z\simeq 2-4$ (e.g. \citealt{teplitz00, pettini01, mclinden11}, 
 and those harboring AGN; \citealt{maschietto08, kuiper11}) and well above the [OIII] luminosity 
the SED derived SFRs would indicate.


At $z \gtrsim 3.3$, both [NII]$\lambda 6584$ and H$\alpha$ fall redward of the 
$K$-band, i.e., at wavelengths where spectroscopy is not yet feasible. In the 
absence of both these nebular emission lines, it is not possible to 
compare the line ratios of the galaxy on the most widely used emission line 
diagnostic, the BPT diagram \citep{bpt81}, which employs optical 
nebular line ratios of [NII]/H$\alpha$ and [OIII]/H$\beta$ to distinguish 
star-forming galaxies, Seyfert 2 galaxies (or AGN NLRs) and low-ionization 
nuclear emission-line region (LINER) galaxies.

Recently, \citet{juneau11,juneau14} introduced a new excitation diagnostic 
that could be used in the absence of [NII] and H$\alpha$ measurements, namely 
the Mass-Excitation diagnostic (MEx). This diagnostic uses the well-known 
correlation between galaxy stellar mass and gas phase metallicity 
\citep{tremonti04} to create an alternate BPT-like diagram with [OIII]/H$\beta$ 
vs. stellar mass. We determine a lower limit to [OIII]/H$\beta$ ratio by 
assuming that H$\beta$ has the same width as [OIII] and a peak flux three 
times the rms measured in the same spectral region, implying 
$\log{(\rm{[OIII]/H\beta)}}\ge 1.27$. Given the stellar mass derived from SED 
modeling, i.e., $\log{(M_{*}/M_{\sun})}=11.49$, the estimated lower limit on 
$\log{(\rm{[OIII]/H\beta)}}$ implies the presence of an AGN in C1-23152, based 
on its position on the MEx diagnostic diagram. We also note that the estimated 
lower limit on $\log{\rm{([OIII]/H\beta)}}$ is a factor of $\sim$4 larger 
compared to the unusually high mean [OIII]/H$\beta$ ratio found by 
\citet{holden14} in a sample LBGs at $z\sim3.5$ with 
$\log{(M_{*}/M_{\sun})}<10.5$ and significantly larger specific SFR (e.g., 
log(sSFR/yr$^{-1}$)$>-9$) compared to C1-23152.

The large values for the line ratio of [OIII]/H$\beta$ are 
inconsistent with stellar photoionization, and indicative of photoionization 
by an AGN. The AGN-interpretation is further confirmed by the 
comparably weak [OII] emission, such that the [OIII]/[OII] ratio 
$\sim 2.2^{+3.6}_{-1.7}$ (measured assuming identical widths and redshift 
for both lines). This line ratio is a sensitive diagnostic of the ionization 
parameter, and efficiently separates Seyfert-like objects from star-forming 
galaxies and LINERS at low redshift \citep{kewley06}. Because we lack any 
constraints on [OI]/H$\alpha$ that could further help separating the AGN from 
star-forming and composite galaxies, at [OIII]/[OII]$>$1 the main ambiguity is 
between Seyferts and low metallicity galaxies with significant ongoing star 
formation \citep{dopita06, nakajima13}. The upper limit on SFR obtained from 
[OII] line luminosity along with the small values of SFRs from the SED 
modeling provides strong evidence that the galaxy is not a star-bursting 
galaxy. The evaluation of C1-23152 using the MEx diagnostic further supports 
the AGN-like nature of the line emissions.

\subsection {Continuum Emission from the AGN}\label{maxAGNcont}
In the previous sections, we found that C1-23152 likely hosts a luminous AGN 
from emission line properties. The AGN continuum emission could potentially 
contribute to the observed SED, biasing the derived stellar population 
parameters. We investigated this by subtracting the AGN contribution from the 
observed SED and by re-fitting the corrected SED. Specifically, we assumed a 
power-law SED for the AGN, with $F_{\nu}\; \propto\; \nu^{\alpha}$. The value 
of $\alpha$ was derived by fitting the rest-frame UV and the MIPS 24$\mu$m 
photometry and spectroscopy for maximal AGN SED contribution, corresponding to 
$\alpha\;\sim\;-1.6$. The maximum AGN contribution is then set by the 
rest-frame UV fluxes in combination with the 24$\mu$m band, and subtracted 
from the observed SED. The resulting SED is finally re-modeled using FAST to 
derive stellar population parameters (listed in Table~\ref{tab-sed}). We find 
that the derived stellar mass is smaller by $\sim 0.05$~dex, resulting in 
$\log{(M_{*}/M_{\sun})}=11.44$ (i.e., $M_{*}\approx 2.8 \times 10^{11}$ 
M$_{\sun}$).

\begin{figure}
\epsscale{1.15}
\plotone{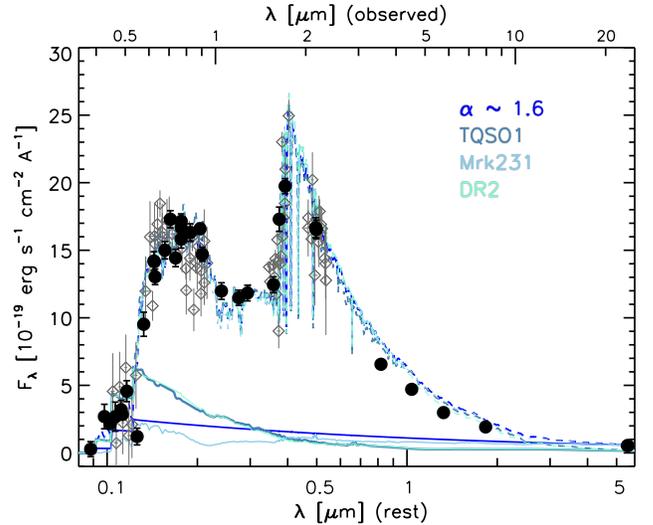}
\caption{Observed SED of C1-23152 from the $U$-band to the MIPS 24$\mu$m; 
filled black circles represent the broad- and medium-band photometry, while 
the gray open diamonds show the binned spectra. The continuous curves show the 
four AGN templates with their maximum contributions constrained by the observed 
24$\mu$m flux and/or the rest-frame UV flux. The dashed curves show the 
combinations of the AGN templates and the best-fit FAST models obtained when 
fitting the AGN-continuum corrected SED of C1-23152. The different colors 
represent the four adopted AGN templates. Templates for Mrk231, TQSO1 and DR2 are 
from \citet{polletta07} and \citet{salvato09}: TQSO1 and DR2 represent high IR luminosity and low 
luminosity QSO templates respectively, with a power law component extended into the UV.
\label{sedwAGN}}
\end{figure}

We also used QSO templates from \citet{polletta07} and \citet{salvato09} to fit the maximum AGN 
continuum contribution in a similar manner as above to constrain the stellar 
population parameters of the galaxy. The templates used include the heavily 
obscured broad absorption line (BAL) QSO, Mrk231, and optically selected QSO 
with different IR to optical flux ratios. The AGN-continuum subtracted SEDs 
were re-modeled using FAST with a Kroupa IMF, exponentially declining SFH and 
BC03 stellar population synthesis models, and the best-fit stellar properties 
are listed in Table~\ref{tab-sed}. Figure~\ref{sedwAGN} shows the observed SED 
of C1-23152 from the $U$-band to the MIPS 24$\mu$m overplotted on the adopted 
AGN templates with their maximum contributions constrained by the observed SED. 
Also plotted are the combinations of the best-fit FAST models of the 
AGN-continuum subtracted SED and the adopted AGN templates. The best fit 
stellar population parameters are again consistent with zero dust attenuation 
and low SFR; the stellar mass of the galaxy is lower on average by 
$\sim\;0.05$ dex, compared to best-fit values modeled using the observed SED. 
The minimum stellar mass allowed by the 3$\sigma$ uncertainties 
is $\log{(M_{*}/M_{\sun})}=11.32$ (i.e., $M_{*}\approx 2.1 \times 10^{11}$ 
M$_{\sun}$), confirming that the observed SED of the galaxy is dominated by 
the stellar light, rather than the AGN emission, and confirming the very large 
stellar mass of C1-23152. 

\subsection{Further Evidence for Hidden Luminous Quasar}

\subsubsection{AGN luminosity}

The AGN-like line ratios and line widths suggest that C1-23152 hosts a 
luminous AGN. If we ignore the contribution from strong ionizing radiation 
fields due to star formation, we can use the [OIII] luminosity as a direct 
proxy for the bolometric luminosity of the quasar. We used the bolometric 
correction ($C_{\rm [OIII]}$) for extinction corrected $L_{\rm [OIII]}$ derived by 
\citet{lamastra09} $L_{\rm bol,AGN}\;\approx\;454\;L_{\rm [OIII]}$ for 
extinction-corrected [OIII] luminosities to find 
$L_{\rm bol}=7.5^{+5.9}_{-3.3}\times 10^{45}$~erg~s$^{-1}$. 
The authors in this study use a sample of type-2 AGN in SDSS with reliable 
$L_{\rm [OIII]}$ and X-ray luminosities ($L_{\rm X}$) to estimate $C_{\rm [OIII]}$ 
combining the observed correlation between $L_{\rm [OIII]}$ and $L_{\rm X}$  
with the X-ray bolometric correction \citep{marconi04}.  We also 
estimated the ionizing luminosity, $L_{\rm ion}$, from the total luminosity in 
narrow lines, $L_{\rm NLR}$. Using $L_{\rm NLR}=3(3L_{\rm [OII]}+1.5L_{\rm [OIII]})$ 
and $L_{\rm ion}=L_{\rm NLR} \times C^{-1}$ and adopting a covering factor of 
$C\sim 10^{-2}$ \citep{rawlingssaunders91}, we find 
$L_{\rm ion } = 14.2^{+6.6}_{-4.5} \times 10^{45}$~erg~s$^{-1}$. Both estimates 
imply the presence of a luminous hidden quasar.  

\subsubsection{Infrared Spectral Energy Distribution}


\begin{figure}
\epsscale{1.15}
\plotone{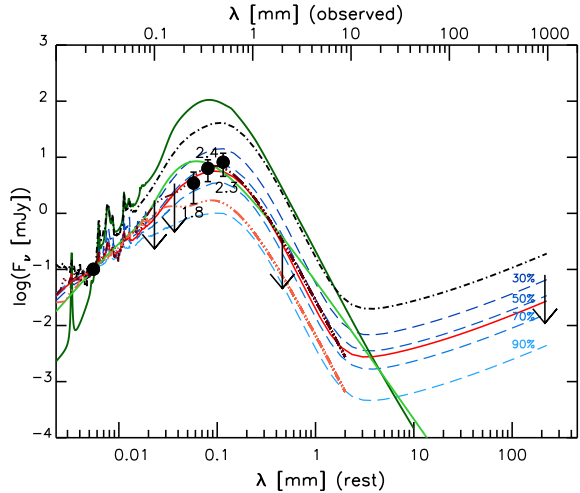}
\caption{IR SED with \emph{Spitzer} MIPS$24\mu$m, \emph{Herschel} PACS $100\mu$m, $160\mu$m, 
SPIRE $250\mu$m, $350\mu$m, $500\mu$m, GISMO 2mm and 320 MHz data from GMRT with IR templates 
fit to $24\mu$m photometry overplotted. \emph{Filled black circles} represent observations 
with detections above $1\sigma$ with the significance of detection indicated by the S/N at each 
point. Photometric points with no detection are indicated as $1\sigma$ upper limits. 
\emph{Light green solid} template represents the best fit obtained using the FIR black-body models 
from \citet{casey12}; \emph{dark green solid} template represents the starburst SED from 
\citet{magdis12}; high-$z$ composite AGN templates with and without silicate features from 
\citet{kirkpatrick12} are plotted in   
\emph{dark red} and \emph{orange triple-dot dashed} lines respectively; \emph{dashed (blue)} templates represent 
the mean SEDs for $\alpha=1,...,2.5$ from the template set of \citealt{dale14} for varying 
degrees of AGN contribution (\emph{dark blue}: 0$\%$ to \emph{light blue}: 100$\%$ AGN 
contribution in uniform $20\%$ steps); \emph{red solid} SED is the 
template from \citet{dale14} which minimizes $\chi^{2}$ and \emph{black dot dashed} SED represents the average 
SED of \citealt{dale02} templates used in \citet{marchesini10}. IR templates with 
only obscured star-formation cannot model the full observed IR SED, an AGN contribution of $>50\%$ 
is required to reproduce the observed IR SED. 
\label{fig-fnu}}
\end{figure}

Figure~\ref{fig-fnu} shows the observed far-infrared (FIR) SED of C1-23152, including 
fluxes in the {\it Spitzer}-MIPS 24$\mu$m (from the NMBS catalog of \citealt{whitaker11}), 
{\it Herschel}-PACS 100$\mu$m, 160$\mu$m and SPIRE 250$\mu$m, 
350$\mu$m, 500$\mu$m, the 2mm Goddard-IRAM Superconducting 2 Millimeter 
Observer (GISMO) on the IRAM 30m telescope (Karim et al. 2015, in 
preparation), and the 320~MHz data from the Giant Meterwave Radio Telescope 
(Karim et al. 2015, in preparation). The PACS and SPIRE photometry was extracted 
from the images of the PACS Evolutionary Probe (PEP) survey \citep{lutz11}, 
the Herschel Multi-tiered Extragalactic Survey (HerMES; \citealt{oliver12}), 
and Herschel-CANDELS (PI: Dickinson) as described in Bedregal et al. 
(2015, in preparation). 
The 1-$\sigma$ detection limits of Herschel photometry are 1.7 and 3.4 mJy for PACS 100$\mu$m \citep{berta11}, 160$\mu$m and 3.2, 2.7, 3.8 mJy for SPIRE 250$\mu$m, 350$\mu$m, 500$\mu$m photometry \citep{oliver12}, respectively.

The source is not robustly detected ($>$3$\sigma$) in 
any of the FIR bands, except in the MIPS 24$\mu$m ($\sim$7$\sigma$); the SPIRE 
fluxes have a S/N$\lesssim$2. The source is also not detected in the Submillimeter 
Common User Bolometric Array-2 
(SCUBA-2; \citealt{holland13}) deep 450$\mu$m and 850$\mu$m observations 
on the James Clerk Maxwell Telescope (JCMT) \citet{casey13}, or in 1.1~mm AzTEC 
Surveys \citep{scott08, aretxaga11}. 

The observed 24$\mu$m band probes the rest-frame $\sim$5.5$\mu$m, 
corresponding to the spectral region of thermal emission from hot dust at the 
redshift of C1-23152. The source of the mid-infrared (MIR) emission was 
investigated by \citet{marchesini10}. Assuming that all the luminosity at 
the observed 24$\mu$m is associated with dust-enshrouded star formation, 
\citet{marchesini10} adopted the mean of 
$\log{L_{\rm IR,\alpha=1,...,2.5}}$ from the template set of \citet{dale02} 
(presented in Fig.~\ref{fig-fnu} as dot-dashed black curve) to estimate a total 
8-1000$\mu$m rest-frame IR luminosity 
$L_{\rm IR}$=1.14$\times$10$^{13}$~L$_{\odot}$, corresponding to a 
SFR$\sim$1260~M$_{\odot}$~yr$^{-1}$ when using the $L_{\rm IR}$-SFR calibration 
adapted for a Kroupa (2001) IMF from \citet{kennicutt98}. Larger values of 
$L_{\rm IR}$ and SFR would be estimated if using the starburst template from 
Magdis et al. (2012; dark green curve in Fig.~\ref{fig-fnu}). In place of the 
\citet{dale02} templates, we also use the calibration by \citet{rujopakarn13} 
to obtain $L_{\rm IR}$=7.45$\times$10$^{12}$~L$_{\odot}$ and 
SFR$\sim$835~M$_{\odot}$~yr$^{-1}$. An independent estimate of $L_{\rm IR}$ 
can be finally derived from the IDL code of \citet{casey12} to fit a modified 
black-body to the FIR detections and upper limits, resulting in 
$L_{\rm IR}$=4.34$\times$10$^{12}$~L$_{\odot}$ and 
SFR$\sim$480~M$_{\odot}$~yr$^{-1}$ (the light green SED in Fig.~\ref{fig-fnu} represents 
this best-fit model). All of these values of SFR are from many hundreds to a 
couple of thousands times larger than the SFRs estimated from SED modeling, 
and a factor of $\sim$30-45 larger than the SFR derived using the 
AGN-corrected [OII] line luminosity. 

The inconsistency of the 24$\mu$m-derived SFR, which assumes that the MIR 
emission originates from dust-enshrouded star formation, compared to the SED- 
and [OII]-derived SFRs provides further evidence that C1-23152 hosts a 
powerful AGN. Moreover, as shown in Figure~\ref{fig-fnu}, no model in which the IR emission 
is due only to obscured star formation can reproduce the observed IR SED. 
Indeed, significant contributions from an obscured AGN is required to match the 
observations. We use the IR template set of \citet{dale14} which 
includes fractional AGN contributions to IR radiation to investigate 
constraints on the source of radiation responsible for FIR detections. For 
visual ease in comparison, we calculate the mean $\log{L_{\rm IR}}$ for 
$\alpha=1,...,2.5$ from this template set for AGN contributions of 30\%, 50\%, 
70\%, and 90\%, scaled to the observed MIPS 24$\mu$m photometry. These mean 
templates are plotted as blue dashed lines in Figure~\ref{fig-fnu} with the contribution 
of AGN in 20\% increments. We fit the detections and the upper limits with the 
individual templates from \citet{dale14}. The template with 60\% AGN 
contribution and $\alpha=1.6875$ yields the minimum $\chi^2$ and is shown in 
Figure~\ref{fig-fnu} as a red solid curve. The IR detections and 1$\sigma$ upper 
limits effectively rule out dust-enshrouded star formation as the only source 
of the observed 24$\mu$m flux, with the AGN contributing $>$60\% to the IR SED 
of C1-23152.

To estimate the bolometric luminosity of the quasar independently of emission 
line luminosities, we used the two high-$z$ composite templates of AGNs from 
\citet{kirkpatrick12}, one with a clear 9.7$\mu$m silicate absorption feature 
and the other with a featureless MIR spectrum. We scaled the AGN templates
from \citet{kirkpatrick12} (plotted in Fig.~\ref{fig-fnu} in dark red and orange 
triple-dot dashed curves for the templates with and without silicate features, 
respectively) to the observed 24$\mu$m (rest-frame 5.5$\mu$m)
luminosity of C1-23152.

Using the integrated luminosities of the SED templates, we find 
$L_{\rm bol,AGN} \approx (8.9 \pm 2.7) \times 10^{45}$~erg~s$^{-1}$ and 
$L_{\rm bol,AGN} \approx (15.2 \pm 6.5) \times 10^{45}$~erg~s$^{-1}$ for the 
featureless AGN template and the AGN template with silicate absorption, 
respectively, in good agreement with the bolometric AGN luminosities estimated 
from the emission line luminosities. 

\begin{figure*}
\epsscale{0.7}
\plotone{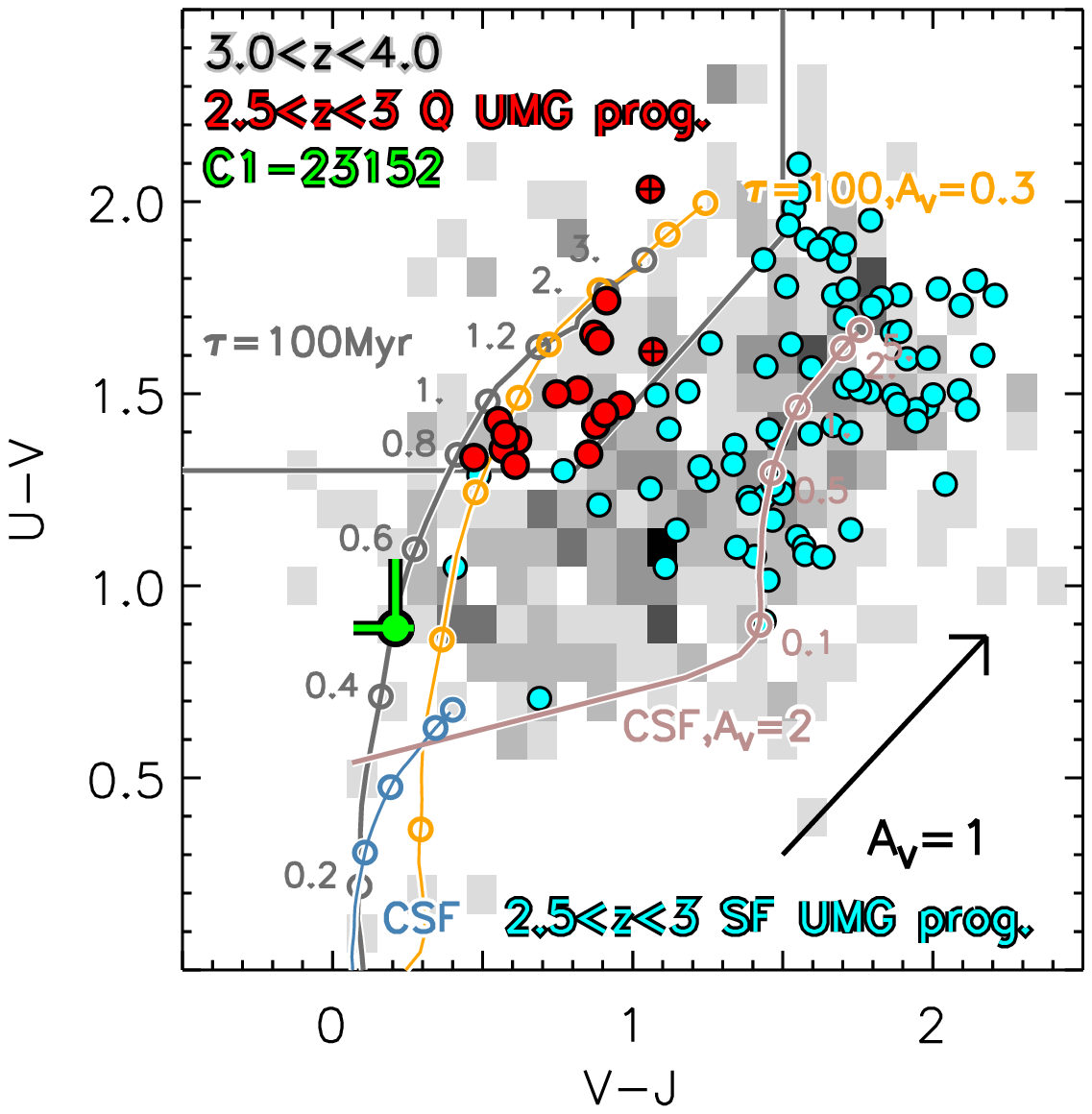}
\caption{Rest frame $U-V$ versus $V-J$ color-color diagram. The 
\emph{grayscale} representation indicates the distributions of all galaxies 
with $M_{\star}>5\times10^{10}~M_{\odot}$ at 
$3.0\;\le\;z\;\le\;4.0$ from the $K_{s}$-selected UltraVISTA catalog 
\citep{muzzin13a}. The cuts used to separate star-forming from quiescent 
galaxies from \citet{muzzin13b} are shown as the solid dark gray lines. The 
\emph{green point} indicates the color for C1-23152, with error bars 
representing the total 1$\sigma$ uncertainties. \emph{Red filled points} and 
\emph{blue filled points} are the quiescent and star forming progenitors, 
respectively, of local UMGs at $2.5\;<\;z\;<\;3.0$ from 
\citet{marchesini14}, with crosses indicating sources with MIPS detection at 
$\ge\;5\;\sigma$. Color evolution tracks of \citet{bc03} models are also 
shown: an exponentially declining SFH with no dust ($\tau=100$ Myr; gray), 
the same exponentially declining SFH with $A_{\rm V}=0.3$ mag ($\tau=100$ Myr; 
orange), a constant SFH with no dust (CSF; light blue), and the same CSF model 
with $A_{\rm V}=2$ mag of extinction (light brown). The empty circles represent 
the model colors at the specified ages (in Gyr). The dust vector indicates an 
extinction of $A_{\rm V}=1$ mag for a \citet{calzetti00} extinction curve. 
\label{fig-UVJ}}
\end{figure*}

\subsubsection{Radio/X-ray Emission}
We have used the publicly available Chandra X-ray data available over the 
COSMOS field \citep{elvis09} to search for X-ray detection. C1-23152 is not 
detected. Using the 3$\sigma$ detection flux limits, we derived the 3$\sigma$
upper limit to the rest frame 2-10~keV luminosity of 
$L_{\rm 2-10keV} \approx 1.9 \times 10^{44}$~erg~s$^{-1}$, assuming a power-law 
photon index of $\Gamma=1.9$ \citep{nandrapounds94}. 

To investigate whether the expected X-ray flux of C1-23152 is below this 
detection limit, we used two methods to calculate the expected X-ray 
luminosity. We first employed the relation between mid-infrared (12$\mu$m) and 
X-ray luminosities observed by \citet{gandhi09} in local Seyferts, assuming the 
two high-$z$ AGN SEDs as above. We found 
$L_{\rm 2-10keV} \approx 9.0 \pm 3.5 \times 10^{44}$~erg~s$^{-1}$ and 
$\approx 9.1 \pm 3.6 \times 10^{44}$~erg~s$^{-1}$ for the featureless and 
silicate AGN templates, respectively.

We also estimated the rest-frame $L_{\rm 2-10keV}$ luminosity using the 
bolometric correction adopted from \citet{hopkins07} again using the 
same two AGN SED templates as above. The resulting X-ray luminosities are 
$L_{\rm 2-10keV} \approx 1.6 \pm 0.5 \times 10^{44}$~erg~s$^{-1}$ and 
$\approx 2.4 \pm 1.1 \times 10^{44}$~erg~s$^{-1}$ for the silicate absorption 
and the featureless templates, respectively. The estimated X-ray luminosities 
using the \citet{gandhi09} relation imply that the non-detection of C1-23152 
in the X-rays would require a Compton-thick nature of the AGN, whereas the 
X-ray luminosities estimated using the bolometric correction from 
\citet{hopkins07} can be compatible with the observational flux limitations. 
We stress that these calibrations are based on low-redshift samples and the 
large uncertainties associated in both estimates prohibit us  
making firmer conclusions about the X-ray properties of C1-23152. 

C1-23152 is also not detected in the publicly available 1.4 GHz VLA radio data
available over the COSMOS field \citep{schinnerer10}. The $3\sigma$ limit for 
the intrinsic $L_{\rm 1.4GHz }$ at the redshift of our target is sufficient to 
rule out a radio-loud AGN (i.e., $\log{(L_{\rm 1.4GHz}[\rm{W~Hz^{-1}}])}$<25; \citealt{schinnerer07}), but 
does not provide any meaningful SFR estimates.
Given the numerous evidence for the (almost) post-starburst state of 
C1-23152, a non-detection is not surprising, as radio-loud AGNs are commonly 
found in hosts with SFR$\sim$300 M$_{\sun}$yr$^{-1}$ at $z\sim$2 
\citep{floyd13}. 

\subsection {UVJ Colors}

The rest-frame $U-V$ versus $V-J$ diagram (hereafter, $UVJ$ diagram) has 
become a popular way to differentiate between star-forming and quiescent 
galaxies, with galaxy populations clearly separated in this color-color space 
\citep{labbe05, wuyts07, williams09, brammer11, whitaker11, patel12,patel13, 
muzzin13b}. Figure~\ref{fig-UVJ} shows the rest-frame $U-V$ and $V-J$ colors of 
C1-23152 (green) overplotted to the distribution in the $UVJ$ diagram of all 
galaxies at $3.0<z<4.0$ from the $K_{\rm S}$-selected UltraVISTA catalog of 
\citet{muzzin13a} (grayscale representation). We calculated the rest frame 
$U-V$ and $V-J$ colors of C1-23152 using EAZY \citep{brammer08}. A Monte Carlo 
approach was used to measure the uncertainties in the $U-V$ and $V-J$ colors. 
Specifically, 1000 photometry catalogs were created by perturbing each flux by 
a Gaussian random number with the standard deviation set by the level of each 
flux error. The simulated catalogs were each fit with EAZY separately, and the 
formal upper and lower limits were obtained in a similar manner as for the 
emission line fits. 

The colors of C1-23152, $U-V=0.89^{+0.18}_{-0.02}$ mag and 
$V-J=0.21^{+0.06}_{-0.14}$ mag, place it formally outside the commonly defined 
quiescent region, with a $U-V$ color intermediate between a star-forming 
galaxy with no dust and a quiescent post-starburst galaxy. 
Figure~\ref{fig-UVJ} also shows color-color evolution tracks for different 
SFHs and dust extinction. The gray and orange tracks represent the evolution 
of an exponentially declining SFH with $\tau=100$~Myr and $A_{\rm V}$=0 and 
0.3~mag, respectively. The blue and light brown tracks represent the color 
evolution of constant star formation history with $A_{\rm V}$=0 and $2.0$ mag, 
respectively. Figure~\ref{fig-UVJ} shows that the derived $U-V$ and $V-J$ 
colors of C1-23152 match the colors predicted for a stellar population 
characterized by an exponentially declining SFH with $\tau=100$~Myr, no dust 
extinction, and an age of $\approx$0.5 Gyr, consistent within the errors with 
the stellar population properties of C1-23152 derived from the SED modeling.

As described in Section~\ref{sec-sedmodel}, we performed SED fits 
for composite SFHs that include an old and young population. 
We used the best-fit parameters obtained for the composite SFH to investigate 
the rest-frame U-V and V-J colors of the underlying old population. For 
the two-component SFH models that reproduce the observed SED reasonably well, 
we found that the rest-frame U-V and V-J colors of the old stellar population 
are only slightly redder, $\langle\Delta$(U-V)$\rangle$~=~$\langle\Delta$(V-J)$\rangle$~$\sim$~0.3~mag. This is 
expected as by construction, composite SFHs will result in the old stellar component 
to be slightly more evolved compared to stellar populations obtained from simple SFHs. 
Nevertheless, the colors obtained of the old stellar components for these
SFHs are still similar to those derived assuming an exponentially declining SFH and are
typical of a galaxy in a post-starburst phase about to enter the quiescent box.
We therefore conclude that the influence of a recent burst does not change the overall 
interpretation of the color evolution of C1-23152.

\subsection {Size} \label{sec-size}

\begin{figure*}
\epsscale{1}
\plotone{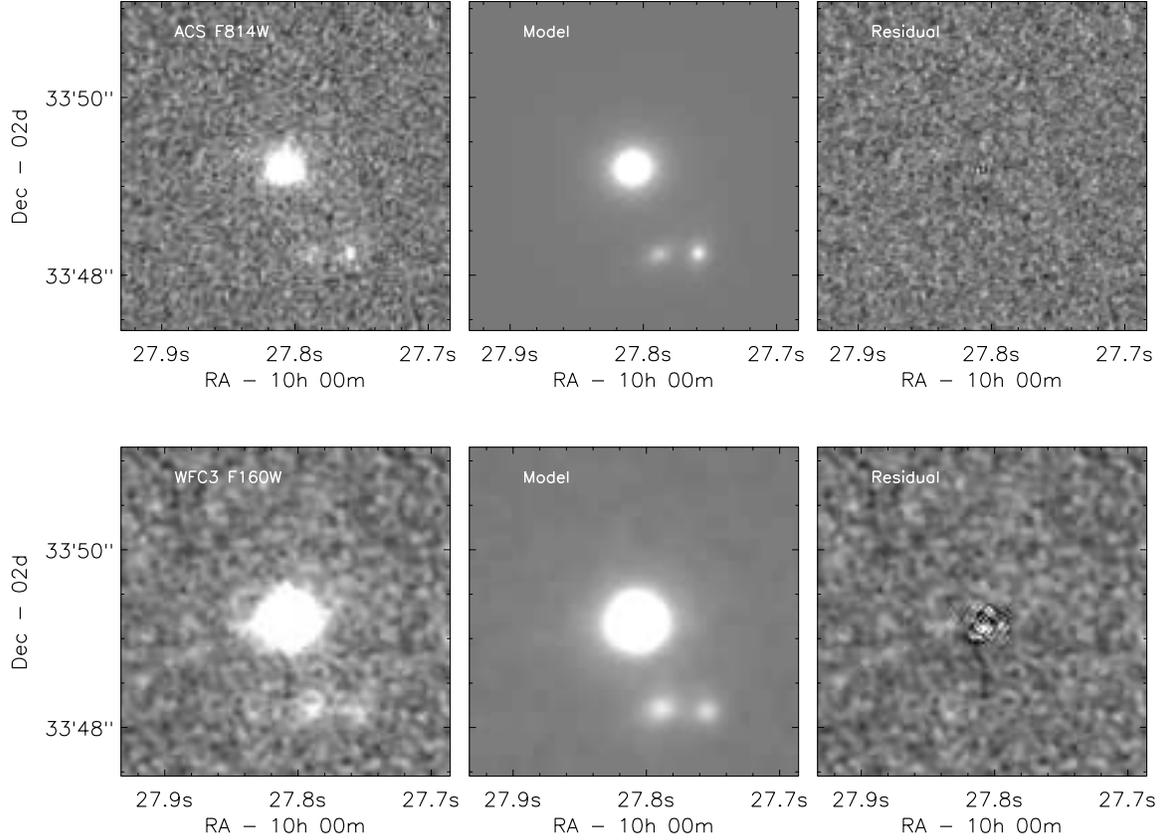}
\caption{ \emph{Left Column:} ACS (top) and WFC3 (bottom) images for C1-23152. The 
galaxy is resolved in both bands. \emph{Central column:} Best fit GALFIT modeling. 
\emph{Right column:} Residual images. The best-fit GALFIT values are $r_{\rm e,circ}=1.45\pm0.15$~kpc, 
$n=5.5\pm0.5$ and $b/a=0.85\pm0.04$ for ACS imaging, and $r_{\rm e,circ}=0.97\pm0.10$~kpc, 
$n=4.4\pm0.4$ and $b/a=0.82\pm0.03$ for WF3 imaging. C1-23152 is more compact (by $\sim50\%$) 
in H$_{\rm 160}$ (rest-frame U-band) than in the ACS I$_{\rm 814}$ (rest-frame UV). 
There is no evidence for a central point-like source in either bands.\label{fig-images}}
\end{figure*}

We used ACS $I_{814}$ and \emph{HST} F$160W$ imaging to measure the structural 
parameters of the galaxy probing the rest frame UV 
($\lambda_{\rm rest}\approx$1870~\AA) and the rest-frame optical 
($\lambda_{\rm rest}\approx$3680~\AA), respectively. Structural parameters were
obtained with GALFIT \citep{peng02}, which provides measurements of the 
S\'ersic index (\emph{n}), effective radius ($r_{e}$), and axis ratio ($b/a$). 
Visual inspection of the HST images showed that C1-23152 is accompanied by 
two fainter objects $\lesssim 2^{\prime \prime}$ south-west of its center. 
Although much fainter than C1-23152 (by $\sim$2.5 and 3.6~mag in the 
H$_{\rm 160}$ band, and $\sim$2.6~mag in the I$_{\rm 814}$ band), these two 
objects could contaminate the light profile of C1-23152, potentially 
introducing systematic effects in the recovery of the structural parameters. 
The measurements were therefore performed fitting the light profiles of the 
three sources simultaneously.

\begin{figure}
\epsscale{1}
\plotone{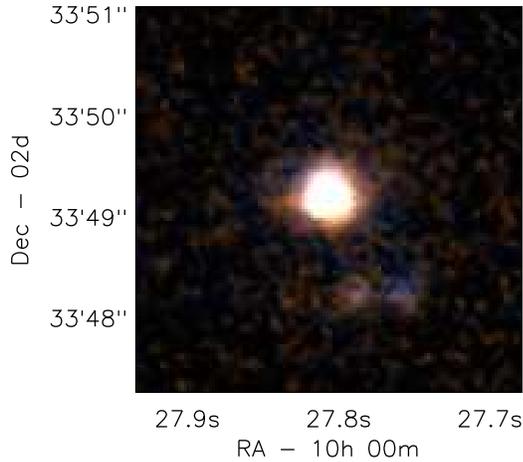}
\caption{Color image obtained from the combination of the HST ACS I$_{\rm 814}$ 
and WFC3 H$_{\rm 160}$ bands centered around C1-23152. C1-23152 shows a smooth 
and round morphology; the two nearby fainter sources are characterized by 
bluer colors. \label{fig-HSTcolor}}
\end{figure}

For each object, a single S\'ersic profile was considered. The effective 
radius was circularized following the relation 
$r_{\rm e,circ}=r_{\rm e} \sqrt{b/a}$. The point spread function (PSF) in each 
HST band was built from a sample of bright and isolated point sources and 
combined together with the IRAF task \url{daophot/psf}. The PSF FWHM for 
the ACS I$_{\rm 814}$ is 0.1$^{\prime \prime}$ and 0.17$^{\prime \prime}$ for the 
WFC3 H$_{\rm 160}$ band . The values recovered for the circularized effective 
radius are $r_{\rm e,circ}=0.195^{\prime \prime}\pm0.021^{\prime \prime}$ and 
$r_{\rm e,circ}=0.130^{\prime \prime}\pm0.013^{\prime \prime}$ for the ACS 
I$_{\rm 814}$ and the WFC3 H$_{\rm 160}$ bands, respectively. The estimated 
values of the S\'ersic index and the axis ratios are $n=5.5\pm0.5$ and 
$n=4.4\pm0.4$, and $b/a=0.85\pm0.04$ and $b/a=0.82\pm0.03$ for the 
I$_{\rm 814}$ and the H$_{\rm 160}$ bands, respectively. Figure \ref{fig-images} 
shows, for each band, a cutout from the original frame centered on C1-23152, 
the GALFIT model, and the residual image after subtracting the model from the 
original frame. As a consistency check, we also run GALFIT masking the two 
neighboring sources and verifying that the morphological parameters obtained 
in this way did not differ sensibly from those obtained including the two 
neighboring sources. Figure~\ref{fig-HSTcolor} shows a color image obtained 
from the combination of the I$_{\rm 814}$ and H$_{\rm 160}$ bands of a cutout 
centered around C1-23152. Whereas C1-23152 shows a smooth and round 
morphology, with no obvious radial color gradient, the two nearby fainter 
sources are clearly bluer. 

Observationally, an AGN could manifest itself as a brighter point source in 
the central region of the galaxy. Visual inspection of the residual images 
from the S\'ersic profile fitting process did not show any evident sign of a 
residual point source. However, its existence was also tested by including a 
point-source component and re-running GALFIT. The resulting flux in each band 
was a value consistent with no flux, causing GALFIT to crash, and confirming 
the result from visual inspection. We finally re-run GALFIT by fixing the 
magnitude of the central point-source component to the values obtained from 
the maximized AGN contributions derived in \S\ref{maxAGNcont}. The 
contributions of the AGN to the total fluxes in the H$_{\rm 160}$ and 
I$_{\rm 814}$ bands is $\sim$8-15\% and $\sim$9-29\%, respectively, depending 
on the AGN template. Fixing the magnitude of the central point-source 
component results in slightly larger effective radii, consistent within 
2$\sigma$ with the values obtained without a central point-source component. 
Specifically, we obtained $r_{\rm e,circ}=0.20^{\prime \prime}-0.23^{\prime \prime}$ 
and $r_{\rm e,circ}=0.14^{\prime \prime}-0.16^{\prime \prime}$ for the ACS 
I$_{\rm 814}$ and the WFC3 H$_{\rm 160}$ bands, respectively, depending on the 
adopted AGN contribution. Whereas the effective radius does not change 
significantly when including a central point-source, the S\'ersic index is much 
more sensitive, and ranges in $n=0.7-3.9$ and $n=3.0-3.6$ for the ACS 
I$_{\rm 814}$ and the WFC3 H$_{\rm 160}$ bands, respectively, depending on the 
adopted AGN contribution. We note however that the inclusion of the central 
point-source component results in increasingly worse modeling of the HST 
images for increasing AGN contributions, especially for the I$_{\rm 814}$ band, 
and that significant positive and negative features appear in the center of 
the galaxy on the residual images output by GALFIT, indicative of 
over-subtraction of the central point-like component.

\begin{figure}
\epsscale{1.15}
\plotone{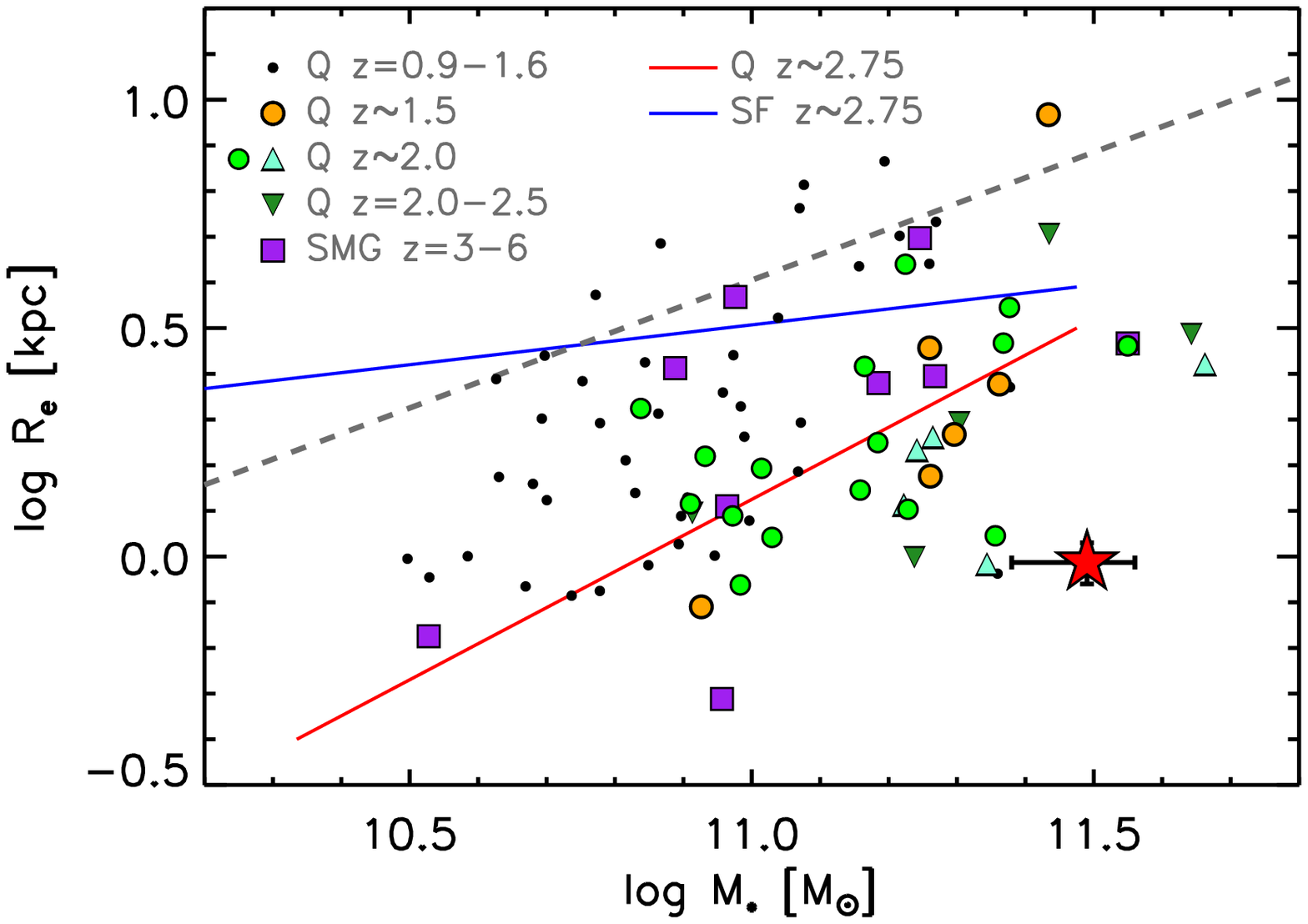}
\caption{Plot of stellar mass versus size. \emph{Red star} indicates the location 
of C1-23152. Spectroscopically confirmed quenched galaxies are indicated in the legend 
with Q and the redshift range of each study (\emph{black points}: \citealt{belli14a}; 
\emph{orange filled circles}: \citealt{bezanson13}; \emph{green filled circles}: \citealt{krogager13}; 
\emph{light blue triangles}: \citealt{vandesande13}; \emph{dark green downward triangles}: \citealt{belli14b}). 
\emph{Purple filled squares} indicate the location of spectroscopically confirmed SMGs from \citet{toft14}.
\emph{Red and blue solid lines} indicate the median mass versus size distribution for quiescent 
and star forming galaxies respectively from \citet{vanderwel14}. \emph{Dark gray dashed line} is the mean 
stellar mass versus size at $z\sim0$ from \citet{shen03}. All measurements are scaled to match a \citet{kroupa01} IMF. 
\label{fig-masssize}}
\end{figure}

Figure~\ref{fig-masssize} shows the stellar mass versus size diagram of spectroscopically 
confirmed quenched galaxies at $z<3$ \citep{bezanson13, krogager13, vandesande13, belli14a, belli14b} 
and spectroscopically confirmed SMGs \citep{toft14}, along with the stellar mass and size 
of C1-23152, highlighting the extreme compactness of this galaxy.
At the redshift of the source, $z_{\rm spec}=3.351$, the measured circularized 
effective radii correspond to linear sizes of $r_{\rm e,circ}=1.45\pm0.15$ 
kpc in the I$_{\rm 814}$ and $r_{\rm e,circ}=0.97\pm0.1$~kpc in the H$_{\rm 160}$ 
band.\footnote{When including a central point-source component with fixed 
magnitude constrained by the maximal contribution from the AGN continuum to 
the observed SED as derived in \S\ref{maxAGNcont}, the sizes increase to 
$r_{\rm e,circ}=1.49-1.73$~kpc in the I$_{\rm 814}$ and $r_{\rm e,circ}=1.05-1.18$ 
kpc in the H$_{\rm 160}$ band, consistent within 2$\sigma$ with the sizes 
obtained without a central point-like component.} Interestingly, the 
effective radius in the rest-frame optical is smaller that the effective 
radius in the rest-frame UV by $\sim$0.5$\pm$0.2~kpc, with a gradient 
$\Delta \log{r_{\rm eff}} / \Delta \log{\lambda}=-0.13\pm0.07$, broadly 
consistent with $\Delta \log{r_{\rm eff}} / \Delta \log{\lambda}=-0.25$ in a 
sample of early-type galaxies at $0<z<2$ found by \citet{vanderwel14}. If we 
consider the circularized effective radius of $\sim$1~kpc derived from the 
H$_{\rm 160}$ band, C1-23152 is a very compact galaxy for its large stellar 
mass, arguably among the most compact very massive galaxies at $z>3$ (see Figure~\ref{fig-masssize}). More 
quantitatively, the size of C1-23152 is a factor of $\sim$2 smaller than the 
median size of galaxies at $2.5<z<3.0$ with similar stellar masses 
($11<\log{M_{*}/M_{\sun}}<11.5$), and it is consistent with the 16\% range in 
the size distribution of this sample \citep{vanderwel14}. We finally note 
that C1-23152 is best modeled by a S\'ersic index $n=4.4\pm0.4$ in the 
H$_{\rm 160}$ band, consistent with a de Vaucouleur's profile ($n=4$), and a 
large axis ratio of $b/a\sim0.8-0.9$. The best-fit structural and stellar 
population parameters for C1-23152 are in line with other studies of 
early-type galaxies at $z>$1 
\citep{franx08,vandokkum11,wuyts11,wake12,bell12,chang13} that have shown that 
sources with high S\'ersic indices and large axis ratios are more likely to be 
quiescent. 


\section{Summary and Discussion} \label{sec-disc}

In this paper we have investigated the extensive (observed-frame) UV-to-NIR 
spectra of a $z_{\rm phot}\approx3.3$, massive galaxy selected from the NMBS, 
namely C1-23152. We confirmed 1) the redshift of the source through the 
analysis of several nebular emission lines, and 2) the very large stellar mass 
of the galaxy by performing stellar population modeling of the spectral energy 
distribution of the galaxy corrected for both emission lines and AGN 
continuum. We also determined through analysis of emission line luminosities 
and line ratio diagnostics that C1-23152 likely hosts a luminous hidden AGN. 

We found that C1-23152 is a very compact ($r_{\rm e}\;\approx$ 1~kpc), very 
massive ($\log{(M_{*}/M_{\sun})}\gtrsim$11.3) galaxy at $z_{\rm spec}$= 3.351 
with a formation redshift of $z\gtrsim 4$ with suppressed star formation, 
about to enter a post-starburst phase (as evidenced from the derived sSFR of 
$\log{(\rm sSFR/\rm yr^{-1})} \sim -11$). We used both BC03 and MA05 stellar 
population models along with the \citet{calzetti00} extinction law and a 
\citet{kroupa01} initial mass function with varying star formation histories 
to model the observed spectral energy distribution. We found that 
independently of SED-modeling assumptions, the best-fit attenuation is 
consistent with zero, the timescale of the burst ($\tau$) is very short 
($\sim$ 50 Myr) and the SFR is low (3$\sigma$ upper limit is a 
7~M$_{\sun}$~yr$^{-1}$). The low SFR inferred from SED fits is inconsistent 
with the SFRs derived from both $L_{\rm [OII]}$ (and only marginally consistent 
when the [OII] luminosity is corrected for AGN contamination) and the observed 
MIPS 24$\mu$m flux if the 24$\mu$m detection is assumed to be due to 
dust-enshrouded star formation (by at least two orders of magnitude). The 
effect of different modeling assumptions change the derived stellar mass by at 
most $\pm$0.08 dex. However, the estimated stellar age 
($\langle t \rangle_{\rm SFR}\;\sim$ 350 Myr when adopting the BC03 models and 
a best-fit solar metallicity) is much more sensitive to the different 
SED-modeling assumptions, with the stellar age a factor of $\sim$2 smaller 
when adopting the MA05 models. We find strong evidence for the presence of a 
luminous hidden AGN ($L_{\rm bol,AGN} \sim 10^{46}$~erg~s$^{-1}$), potentially 
responsible for the quenching of the star formation. We employed the observed 
relation between MIR and X-ray luminosities and bolometric corrections using 
high-$z$ AGN templates to estimate the rest-frame $L_{\rm 2-10keV}$ but cannot 
conclude with certainty whether the non-detection of C1-23152 in the X-ray is 
due to flux limitations of the data or a Compton-thick nature of the AGN.

Independent constraints on the mass of C1-23152 can be obtained from its 
kinematics. Using the velocity dispersion derived from the [OIII] line 
emission, we can attempt to estimate the dynamical mass of the galaxy using 
$M_{\rm dyn}=C\sigma^{2}r_{\rm e}$, where $C$ is a constant that depends on 
the structure of the galaxy and other parameters, and varies between 
$\log{C}=5.87$ (the value resulting in $M_{\rm dyn}\sim M_{\odot}$ for galaxies 
in the SDSS; \citealt{franx08}) and $\log{C}=6.07$ (derived from kinematic 
data of local early-type galaxies; \citealt{vandokkum03}). Using the 
measurements of the effective radius from the H$_{\rm 160}$ and the 
I$_{\rm 814}$ bands, we estimated a dynamical mass ranging from 
$\log{(M_{\rm dyn}/M_{\odot})}=10.91\pm0.24$ (adopting $\log{C}=5.87$ and 
$r_{\rm e}$=0.97 kpc) to $\log{(M_{\rm dyn}/M_{\odot})}=11.28\pm0.24$ (adopting 
$\log{C}=6.07$ and $r_{\rm e}$=1.45 kpc), marginally consistent with stellar 
mass derived from the SED modeling. We note however that this calculation 
assumes that the [OIII] line is a reasonable probe of the overall gravitational 
potential of the galaxy. Whereas this assumption is plausible if the [OIII] 
emission originates from star formation and its line width is integrated over 
the entire galaxy, this does not appear to be the case for C1-23152, as the 
[OIII] emission appears to be mostly produced by the AGN narrow line regions. 
Velocity dispersion measurements from stellar absorption lines are therefore 
needed to robustly derive the dynamical mass of C1-23152. 

Our findings for the properties of this galaxy fit well into the scenario for 
the formation of the progenitors of local giant elliptical galaxies proposed 
by \citet{marchesini14}. \citet{marchesini14} studied the evolution of the 
properties of the progenitors of local UMGs in the last 11.4 Gyr since $z=3$, 
finding that at $2.5<z<3.0$ a small fraction ($\sim15\%$) of the progenitors 
are already quiescent, implying that the assembly of the very massive end of 
the local quiescent red-sequence population must have started at $z>3$. The 
stellar population properties (i.e. stellar age, star formation rate, rest 
frame colors) of C1-23152 are qualitatively and quantitatively consistent with 
it being the progenitor of the already quiescent progenitors of local UMGs at 
$z=2.75$, as clearly evident in Figure~\ref{fig-UVJ}, which shows the relative 
position in the $UVJ$ diagram of C1-23152 and the quiescent progenitors at 
$2.5<z<3.0$ of local UMGs from \citet{marchesini14} (red points in 
Figure~\ref{fig-UVJ} ). Furthermore, Figure~\ref{fig-UVJ} clearly shows that 
the difference in age between C1-23152 and the quiescent progenitors at 
$2.5<z<3.0$ of local UMGs is quantitatively consistent with the cosmic time 
between $z=3.35$ and $z \approx 2.5-3$. 

The large stellar surface densities observed in C1-23152 and other massive 
galaxies at $z=2-3$ imply they must have formed via rapid, highly dissipative 
events at earlier times to account for the relatively small half-light radii 
and high S\'ersic index $n$ \citep{weinzirl11}. Although the exact physical 
picture underlying dissipative collapse mechanisms is uncertain (e.g., 
mergers, tidal interactions, \citealt{ks06, hopkins08, bournaud11, perret14}; gas instabilities in 
turbulent diffuse disks, \citealt{keres05, keres09, dekel09, ricciardelli10, dekel14}, 
cold gas accretion from intergalactic medium (IGM), 
\citealt{sales12, johansson12}), there is evidence that the 
radial gas inflows lead to centrally-concentrated starbursts 
\citep{mihos94,ellison10,scudder12}, along with fueling the central 
super-massive black holes (SMBH). The subsequent energetic feedback from the 
active SMBH may be sufficient to regulate and quench star formation by 
preventing gas cooling \citep{kauffmann00, croton06, hopkins08, vandevoort11}.
Given the correlations between black hole mass, bulge mass, and $n$ 
\citep{peng06,graham07,gultekin09}, our findings that C1-23152 is a very 
massive and compact galaxy with high $n$ undergoing a post-starburst phase 
with evidence for a hidden luminous quasar support this scenario. The 
relatively short lifetime of the quenching phase \citep{mendel13} also 
justifies the low likelihood of observing galaxies in the same transitional 
phase as C1-23152.

Simulations have shown that the formation of compact stellar systems must 
involve highly dissipational mechanisms on short timescales, consistent 
with stellar archeology studies of local massive ellipticals \citet{thomas05}.
Gas-rich major mergers at high redshift serve as plausible examples for such a 
scenario \citep{wuyts10}. As gas is driven to the center, a massive nuclear 
starburst is ignited followed by a AGN/QSO phase that halts star formation, 
and leaving a compact stellar component. SMGs have been proposed as candidates 
to the above scenario 
\citep{toft07, cimatti08, schinnerer08, michalowski10, smolcic11, toft14}.
While the mean size \citep{tacconi06} and dynamical masses \citep{ivison11} of 
SMGs measured through molecular lines have been found to be consistent with 
the properties of quiescent galaxies at $z\;\sim$ 2 (e.g., \citealt{krogager13, toft12}), 
there remains the challenge of robustly constraining the stellar masses of these systems
\citep{michalowski12}.

Recently, \citet{gilli14} presented the study of an ultraluminous infrared 
galaxy at $z = 4.75$ in the Chandra Deep Field South, namely XID403. This 
source is found to be compact ($r \sim 0.9$~kpc), with a stellar mass 
estimated between $<3 \times 10^{10}$~M$_{\sun}$ \citep{debreuck14} and 
$\sim 7 \times 10^{10}$~M$_{\sun}$ \citep{gilli14}, depending on the amount 
of AGN contamination assumed in the rest-frame UV-to-NIR, and harboring a 
Compton-thick QSO. The SFR derived by fitting the far-IR SED is 
$\sim650 M_{\sun}\rm yr^{-1}$ (assuming \citealt{kroupa01} IMF), whereas fitting 
the optical/NIR SED results in $\rm SFR \sim 180$~M$_{\sun}$~yr$^{-1}$ 
\citep{gilli14}. The inferred stellar mass, size, gas depletion timescale 
($\sim 10^{7}$~yr), and SFR of XID403 are consistent with it likely being the 
progenitor of the compact, quiescent, massive galaxies at $z<3$. Our findings 
indicate that C1-23152 represents the transitioning phase between XID403 and 
the compact, massive, quiescent galaxies at $z<3$. Its quenched star-formation 
activity while still harboring a hidden luminous AGN makes C1-23152 a natural 
descendent of XID403. Furthermore, \citet{straatman14} studying massive, quiescent $z\sim4$ 
galaxies have shown that based on number density considerations, the majority of the star
formation in their progenitors could have been obscured by dust. With an intense burst of star formation in the recent 
past of C1-23152, one may expect a significant amount of dust in C1-23152 as 
produced by supernovae. The upper limits on the amount of dust obscuration 
resulting from the SED modeling ($A_{\rm V}<0.4$ mag) seems to imply that the 
mechanism responsible for the quenching of star formation in C1-23152 may also 
be responsible for the destruction of most of the dust produced in the intense 
starburst that produced the very large stellar mass of C1-23152. However, in order
to form conclusive and convincing evolutionary paths for the progenitors of post-starburst
massive galaxies at $z\sim3-4$, it is critical to investigate such scenarios by directly 
selecting progenitors by the use of some tracer, such as fixed number density or abundance matching, 
\citep{behroozi13, marchesini14}, or fixed velocity dispersion \citep{bezanson12, belli14a,belli14b}.
In particular, finding clear and direct evidence linking C1-23152 and 
compact, dusty star-forming galaxies at higher redshifts (such as the SMG studied in \citealt{gilli14}, 
see also \citealt{nelson14}, or $z>3$ analogues of SF progenitors studied in \citealt{patel13, barro14, stefanon14, williams14}) 
would result in important constraints on 
timescales of dust formation and destruction.

A few insights can be derived regarding the evolution of C1-23152 
and the progenitors of local UMGs at $2.5<z<3.0$ when examining 
Figure~\ref{fig-UVJ}. The SF progenitors are predominantly found to 
be red in both U-V and V-J colors, indicating that they are dust-obscured. 
Following the color evolution of local UMG progenitors in \citet{marchesini14} 
to $z\sim1.25$, when all SF progenitors are quenched, the region typically 
populated by post-starburst galaxies in the U-V versus V-J diagram 
is not re-populated, indicating that SF progenitors at $z<3$ appear to 
quench without losing dust, or quenching occurs significantly more rapidly 
than the destruction or removal of dust. The fact that there also exists a clear population of
post-starburst galaxies at $z>3$ with properties similar to C1-23152 (e.g., overall SED shape, 
SFR and dust extinction) with short star-formation time scales and little 
to moderate dust extinction implies that there are multiple
paths for star-forming galaxies to join the quiescent population, characterized 
by different relative timescales responsible for quenching and dust 
removal.

\citet{wellons14}, examining galaxies with properties similar to those of compact, 
massive quiescent galaxies observed at $z\sim2$ with the Illustris cosmological simulation 
find that there are multiple pathways to their formation. The dominant mechanisms 
responsible for the formation of this compact population are found to be either assembling 
at early times or through experiencing central starburst. Galaxies that assemble at
early cosmic epochs when the gas density of the universe is very dense are formed 
compact and remain so, whereas galaxies that experience a central starburst (usually
triggered through gas-rich mergers) become compact at later times (z=2-4). 
While it is possible that both mechanisms play a role to a degree in the formation 
and evolution of galaxies similar to C1-23152, it is essential 
to disentangle the more dominant mechanism. Observationally, comparing 
the sizes of galaxies at different wavelengths can serve as a test 
(with those experiencing central starbursts having smaller sizes at shorter wavelengths 
and vice versa). Investigating the observational effects 
of dust for the different formation mechanisms in simulations is crucial, 
as it will provide insight into the color evolution of compact massive galaxies,
perhaps explaining the multiple quenching modes of UMG progenitors discussed above.

Spectroscopic observations in both the optical and in the NIR of other 
candidates of very massive galaxies at $z > 3$ are of vital importance to 
confirm their redshifts, to better characterize the properties of their stellar 
populations, to distinguish between dust-enshrouded star formation and 
highly-obscured AGN activity, and to determine emission line and AGN 
contamination to the SEDs and mass-to-light ratios. High S/N rest-frame 
optical spectroscopy (especially redward of $\lambda_{\rm rest} \sim 5000$\AA) 
is needed to constrain star formation histories, SFRs, and metallicities, and 
to derive dynamical masses from stellar absorption lines for candidates of very 
massive galaxies in the first 2 Gyr of cosmic history (i.e., $z>3$). NIRSpec 
on the {\emph {James Webb Space Telescope}} will provide the instrumental 
capabilities and sensitivities to successfully target stellar-mass complete 
samples of these candidates and to comprehensively understand the population 
of monster galaxies in the early universe. Observations with the 
{\emph {Atacama Large Millimeter Array}} will also be crucial in 
discriminating between dust-enshrouded star-formation and obscured AGN 
activity as well as providing an independent estimate of the dynamical masses 
of the most massive galaxies through measurements of the kinematics in these 
systems. Finally, current X-ray data are inconclusive as to whether the 
non-detection of C1-23152 in the X-ray is due to observational flux 
limitations or a Compton-thick nature of the AGN hosted in C1-23152. In order 
to constrain the column density of X-ray absorbing matter, and hence whether 
the AGN is Compton-thick, observations in both soft and hard bands are 
necessary. {\emph {The Nuclear Spectroscopic Telescope Array Mission}} 
(NuSTAR; \citealt{harrison13}) will provide an opportunity to attempt 
addressing this issue.


\acknowledgments
We thank Marjin Franx for valuable comments. ZCM gratefully acknowledges support from the John F. Burlingame and the 
Kathryn McCarthy Graduate Fellowships in Physics at Tufts University. DM and ZCM 
acknowledge the support of the Research Corporation for Science Advancement's 
Cottrell Scholarship, the support from Tufts University Mellon Research 
Fellowship in Arts and Sciences, support from the programs HST-GO-12990 
and HST-GO-12177.16, provided by NASA through a grant from the Space Telescope 
Science Institute, which is operated by the Association of Universities for 
Research in Astronomy, Incorporated, under the NASA contract NAS5-26555 and support 
by the National Aeronautics and Space Administration under Grant NNX13AH38G issued 
through the 12-ADAP12-0020 program. MS acknowledges support from the programs HST-GO-12286.11 and 12060.95. AM 
acknowledges support from the Netherlands Foundation for Research (NWO) 
Spinoza grant. AFS acknowledges support from the Spanish Ministry for Economy 
and Competitiveness and FEDER funds through grants AYA2010-22111-C03-02 and 
AYA2013-48623-C2-2, and Generalitat Valenciana projects Prometeo 2009/064 and 
PROMETEOII/2014/060. AK acknowledges support by the Collaborative Research Council 
956, sub-project A1, funded by the Deutsche Forschungsgemeinschaft (DFG).
The Dark Cosmology Centre is 
funded by the Danish National Research Foundation. This study is partially 
based on data products from observations made with ESO Telescopes at the La 
Silla Paranal Observatory under programme ID 087.A-0514. Some of the data 
presented in this study were obtained at the W.M. Keck Observatory, which is 
operated as a scientific partnership among the California Institute of 
Technology, the University of California and the National Aeronautics and Space 
Administration. The Observatory was made possible by the generous financial 
support of the W.M. Keck Foundation. Keck telescope time was granted by NOAO, 
through the Telescope System Instrumentation Program (TSIP). TSIP is funded by 
NSF. The authors wish to recognize and acknowledge the very significant 
cultural role and reverence that the summit of Mauna Kea has always had within 
the indigenous Hawaiian community. We are most fortunate to have the 
opportunity to conduct observations from this mountain. Also based on 
observations made with the Gran Telescopio Canarias (GTC), installed in the 
Spanish Observatorio del Roque de los Muchachos of the Instituto de 
Astrof\'isica de Canarias, in the island of La Palma. This study makes use of 
data from the NEWFIRM Medium-Band Survey, a multi-wavelength survey conducted 
with the NEWFIRM instrument at the KPNO, supported in part by the NSF and 
NASA. We thank the NMBS and COSMOS collaborations for making their catalogs 
publicly available. This work is also partially based on observations taken by 
the 3D-HST Treasury Program (GO 12177 and 12328) with the NASA/ESA HST, which 
is operated by the Association of Universities for Research in Astronomy, 
Inc., under NASA contract NAS5-26555. IRAM is supported by INSU/CNRS (France), 
MPG (Germany), and IGN (Spain). This work was supported through NSF ATI grants 
1020981 and 1106284.


\end{document}